\documentclass[longauth]{aa}  

\usepackage{graphicx}
\usepackage{txfonts}
\begin{document}

   \title{The GAPS programme at TNG. LVII. TOI-5076b: A warm sub-Neptune planet orbiting a thin-to-thick-disk transition star in a wide binary system}


   \author{M. Montalto\inst{1},
           N. Greco\inst{1,2,3},
           K. Biazzo\inst{11},
           S. Desidera\inst{10},
           G. Andreuzzi\inst{16,11},
           A. Bieryla\inst{22,23},           
           A. Bignamini\inst{15}, 
           A.~S.~Bonomo\inst{5},
           C.~Brice\~{n}o\inst{7},            
           L. Cabona\inst{10},
           R. Cosentino\inst{16},
           M.~Damasso\inst{5},  
           A. Fiorenzano\inst{16},
           W. Fong\inst{20},
           B. Goeke\inst{20},
           K.~M.~Hesse\inst{20},
           V.~B.~Kostov\inst{24},
           A. F. Lanza\inst{1},
           D. W. Latham\inst{22},
           N.~Law\inst{8},       
           L. Mancini\inst{4,5,12,13}, 
           A. Maggio\inst{17},
           M.~Molinaro\inst{15},
           A.~W.~Mann\inst{8},
           G.~Mantovan\inst{14},
           L.~Naponiello\inst{4,5},
           D.~Nardiello\inst{14,10},        
           V.~Nascimbeni\inst{10},          
           I.~Pagano\inst{1},
           M.~Pedani\inst{16},
           B. S. Safonov\inst{9},
           G. Scandariato\inst{1},
           S.~Seager\inst{19,20,21},
           V.~Singh\inst{1},
           A.~Sozzetti\inst{5},   
           I. A. Strakhov\inst{9},
           J. N. Winn\inst{18},
           C.~Ziegler\inst{6},
           T.~Zingales\inst{14,10}       
          }

   \institute{
              INAF - Osservatorio Astrofisico di
              Catania, Via Santa Sofia 78, I-95123 Catania, Italy\\
              \email{marco.montalto@inaf.it}
         \and
              Dipartimento di Fisica e Astronomia "Ettore Majorana", Universit\`a di Catania, Via S. Sofia 64, I-95123 Catania, Italy
          \and 
              Scuola Superiore di Catania, Università di Catania, Via Valdisavoia 9, I-95123 Catania, Italy
          \and
              Department of Physics, University of Rome ``Tor Vergata'', Via della Ricerca Scientifica 1, 00133, Rome, Italy
          \and    
              INAF - Osservatorio Astrofisico di Torino, 
             via Osservatorio 20, 10025 Pino Torinese, Italy  
          \and
              Department of Physics, Engineering and Astronomy, Stephen F. Austin State University, 1936 North St, Nacogdoches, TX 75962, USA
          \and
              Cerro Tololo Inter-American Observatory, Casilla 603, La Serena, Chile
          \and
              Department of Physics and Astronomy, The University of North Carolina at Chapel Hill, Chapel Hill, NC 27599-3255, USA        
          \and
              Sternberg Astronomical Institute of Lomonosov Moscow State University, Moscow, 119234 Russia
          \and
             INAF – Osservatorio Astronomico di Padova, Vicolo dell’Osservatorio 5, 35122 Padova, Italy 
          \and
             INAF – Osservatorio Astronomico di Roma, Via Frascati 33, 00040 Monte Porzio Catone (RM), Italy     
          \and
             Max Planck Institute for Astronomy, K\"{o}nigstuhl 17, D-69117 Heidelberg, Germany
           \and
             International Institute for Advanced Scientific Studies (IIASS), Via G. Pellegrino 19, I-84019 Vietri sul Mare (SA), Italy
           \and
             Dipartimento di Fisica e Astronomia “Galileo Galilei”, Universit\`a
             degli Studi di Padova, Vicolo dell’Osservatorio 3, 35122 Padova,
             Italy 
             \and
              INAF – Osservatorio Astronomico di Trieste, via Tiepolo 11, 34143 Trieste      
            \and
              Fundación Galileo Galilei–INAF, Rambla José Ana Fernandez Pérez 7, 38712, Breña Baja, TF, Spain
              \and
              INAF - Osservatorio Astronomico di Palermo "G.S. Vaiana", Piazza del Parlamento 1, 90134 Palermo, Italy
              \and
              Department of Astrophysical Sciences, Princeton University, Princeton, NJ 08544, USA
              \and
              Department of Earth, Atmospheric and Planetary Sciences, Massachusetts Institute of Technology, 77
              Massachusetts Avenue, Cambridge, MA 02139, USA. 
              \and
              Department of Physics and Kavli Institute for Astrophysics and Space Research, Massachusetts Institute of Technology, Cambridge, MA 02139, USA
              \and
              Department of Aeronautics and
              Astronautics, Massachusetts Institute of Technology, 77 Massachusetts Avenue, Cambridge, MA 02139,
              USA.
              \and
              Center for Astrophysics | Harvard \& Smithsonian, 60 Garden Street, Cambridge, MA 02138, USA
              \and
              University of Southern Queensland, Centre for Astrophysics, West Street, Toowoomba, QLD 4350 Australia
              \and
              NASA Goddard Space Flight Center, Greenbelt, MD 20771, USA
        }
                          
   \date{}

 
  \abstract
   {}
   {
    We report the confirmation of a new transiting exoplanet orbiting the star TOI-5076. 
   }
   {
    We present our vetting procedure and follow-up observations which led to the confirmation of the exoplanet TOI-5076b. In particular, we employed high-precision {\it TESS} photometry, high-angular-resolution imaging from several telescopes, and high-precision radial velocities from HARPS-N.  
   }
   {
    From the HARPS-N spectroscopy, we determined the spectroscopic parameters of the host star: T$\rm_{eff}$=(5070$\pm$143) K, log~g=(4.6$\pm$0.3), [Fe/H]=(+0.20$\pm$0.08), and [$\alpha$/Fe]=0.05$\pm$0.06. The transiting planet is a warm sub-Neptune with a mass m$\rm_p=$(16$\pm$2) M$\rm_{\oplus}$, a radius r$\rm_p=$(3.2$\pm$0.1)~R$\rm_{\oplus}$ yielding a density $\rho_p$=(2.8$\pm$0.5) g cm$^{-3}$. It revolves around its star approximately every 23.445 days. 
   }
   {
    The host star is a metal-rich, K2V dwarf, located at about 82 pc from the Sun with a radius of R$_{\star}$=(0.78$\pm$0.01) R$_{\odot}$ and a mass of M$_{\star}$=(0.80$\pm$0.07) M$_{\odot}$. It forms a common proper motion pair with an M-dwarf companion star located at a projected separation of 2178 au. The chemical analysis of the host-star and the Galactic-space velocities indicate that TOI-5076 belongs to the old population of thin-to-thick-disk transition stars.
    The density of TOI-5076b suggests the presence of a large fraction by volume of volatiles overlying a massive core. 
    We found that a circular orbit solution is  marginally favored with respect to an eccentric orbit solution for TOI-5076b. 
   }
   

\titlerunning{TOI-5076b}

\authorrunning{Montalto et al.}

\maketitle


%

\section{Introduction}

More than half of Sun-like stars host planets with radii in between the radius of the Earth and the one of Neptune \citep{howard2012,fressin2013,petigura2018,zhu2021}.
These sub-Neptune-sized objects (r$\rm_p<$4$\rm\,R_{\oplus}$) do not have analogs in our own Solar System and represent a new class of planets identified by exoplanet surveys.  Observations from the NASA \textit{Kepler} mission have demonstrated that the radius distribution of these objects presents a bimodality with a relative paucity of planets with radii between 1.5 R$_{\oplus}$ and 2 R$_{\oplus}$ \citep{fulton2017,fulton2018,VanEylen2018}. This bimodality hints at some mechanisms that may have sculpted the observed radii distribution. 
Planets below the gap are currently interpreted as stripped cores that lost their atmospheres by means of certain physical processes such as photoevaporation \citep{owen2013} or core-powered mass loss \citep{ginzburg2018}, while planets above the radius gap were able to retain their atmospheres or are water worlds \citep{zeng2019}. Increasing the sample of sub-Neptunes with precisely measured masses, radii, and densities found around different physical environments is important when it comes to clarifying the origins of this class of exoplanets.   

It is still debated whether the birthplace of planets within the Milky Way plays an important role with regard to their orbital and physical properties. In the solar neighborhood, stars belong to three main populations: the thin disk, the thick disk, and the halo \citep{gilmore1983}. Thick-disk stars are considered to be older, kinematically hotter, metal-poorer, and more $\alpha$-enhanced than thin-disk stars \citep[e.g., ][]{gilmore1989,bensby2005,adibekyan2012a,recioblanco2014}. \citet{santos2017} suggested that planets belonging to different stellar populations, with different abundance ratios, may originate from planetary building blocks that have significantly different composition trends. For example, metal-poor thick-disk stars are expected to produce building blocks with much lower iron-mass fractions (which may imply smaller core masses) but higher water-mass fractions than thin-disk stars. 

\citet{bashi2020} provided first estimates of the occurrence rates of small close-in planets (2-100 days, 1-20$\rm\, M_{\oplus}$) for FGK
dwarfs in the solar neighborhood thin and thick disks of the Galaxy. They found that the occurrence rates for planets in the two populations are, overall, comparable with the fraction of stars with planets F$\rm_h\,=$ 0.23$_{-0.03}^{+0.04}$ and the average number of planets per star 
n$\rm_p\,=$ 0.36$\pm$0.05. In the metal-poor regime ([Fe/H]$<$-0.25), 
however, \citet{bashi2020} found a significantly larger occurrence rate of small close-in planets around high-$\alpha$ (thick disk) stars than around low-$\alpha$ (thin-disk) stars, while in 
the iron-rich region there seems to be almost no difference between the occurrence rates around thin- and thick-disk stars, though the sample of metal-rich, $\alpha$-enhanced planet hosts is still too small to make any definite conclusion.

\begin{figure}
\includegraphics[width=\columnwidth]{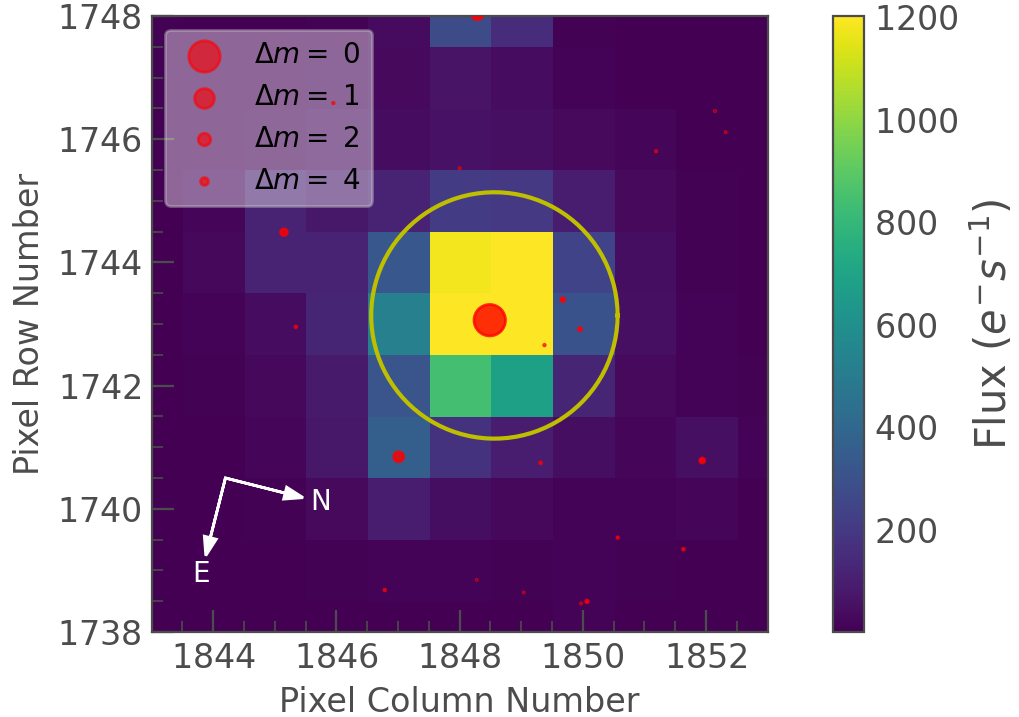}
\caption{
{\it TESS} target pixel file (TPF) showing the position of TOI-5076  relative to the cadence 175714 of sector~42. The image represents an area of 
$\sim$3.5 square arcmin around the target. Sources in {\it Gaia} DR3 are represented by the red dots, scaled inversely proportionally to their difference of apparent $G$-band magnitude with respect to the target and corrected for proper motion to the epoch of the TPF. The yellow circle shows our adopted photometric aperture.
}
\label{fig:tesscut}
\end{figure}

In this work, we report the confirmation of TOI-5076b, a warm sub-Neptune planet orbiting a nearby K2V star member of a binary system and part of the metal-rich, $\alpha$-enhanced population of neighbor stars, with kinematics and chemistry that make it consistent with old, thin-to-thick-disk transition stars. These properties may provide important clues concerning the Galactic formation history of this star and of its planetary system \citep{adibekyan2011}. This is also relevant considering that most transiting exoplanets have been discovered so far around thin-disk stars \citep[e.g., ][]{Biazzo2022}.
Table~\ref{tab:stellar_properties} summarizes some properties of the target star. 

This transiting planet was found using the data provided by the Transiting Exoplanet Survey satellite ({\it TESS}) \citep{ricker2015}. The candidate exoplanet is reported in the ExoFOP database\footnote{\url{https://exofop.ipac.caltech.edu/tess/}}. 
{\it TESS} is a mission that was launched in April 2018 and observes the full sky searching for transiting planets around bright stars. 
The  {\it Global Architecture of Planetary Systems} 
\citep[GAPS,][]{naponiello2022} 
collaboration followed up this target with the 
High Accuracy Radial velocity Planet Searcher for the Northern hemisphere (HARPS-N) spectrograph \citep{cosentino2012}. 
Additional high-angular resolution imaging observations from several observatories have been acquired to further investigate the target.
We analyzed photometric and spectroscopic data simultaneously to infer the properties of the transiting body and of its host star.

In Sect.~\ref{sec:observations}, we describe the photometric, high-angular-resolution imaging and spectroscopic observations we acquired. In Sect.~\ref{sec:data_analysis}, we describe our reduction procedure. In Sect.~\ref{sec:vetting}, we describe our vetting procedure of the planetary candidate. We describe the properties of the host star in Sect.~\ref{sec:properties_of_the_host_star}
 and the planetary properties in Sect.~\ref{sec:planetary_parameters}. 
In Sect.~\ref{sec:discussion}, we discuss our results, and in Sect.~\ref{sec:conclusions} we conclude our analysis.

\begin{table}
        \centering
        \caption{Identifiers, astrometric and photometric measurements, 
             and astrophysical properties of the host star.}
        \label{tab:stellar_properties}
        \begin{tabular}{lcc} 
		\hline
		Parameter & Value & Source \\
		\hline
	    {\it Gaia} & 55752218851048320 & {\it Gaia} DR3$^{1}$ \\
	    TYC          & TYC 1237-472-1 & Simbad \\
	    2MASS        & J03220230+1714240 & Simbad \\
	    TIC          & 303432813 & TIC v8.2$^{2}$ \\
	    TOI          & 5076 & ExoFOP \\
	    $\alpha$(J2016) & 03:22:2.506 & {\it Gaia} DR3 \\
	    $\delta$(J2016) & +17:14:21.07 & {\it Gaia} DR3 \\
	    $\pi$ (mas) & 12.08$\pm$0.01 & {\it Gaia} DR3 \\
	    $\mu_{\alpha}$ (mas yr$^{-1}$) & 168.78$\pm$0.02 & {\it Gaia} DR3 \\
	    $\mu_{\delta}$ (mas yr$^{-1}$) & -176.01$\pm$0.01 & {\it Gaia} DR3 \\
        Distance (pc)      &   82.8$\pm$0.1 & This work\\
	    {\it TESS}     & 9.991$\pm$0.006 & TIC v8.2\\
	    {\it G}        & 10.5928$\pm$0.0002 & {\it Gaia} DR3 \\
	    {\it G$_{BP}$} & 11.1036$\pm$0.0006 & {\it Gaia} DR3 \\
	    {\it G$_{RP}$} & 9.9310$\pm$0.0004 & {\it Gaia} DR3 \\
	    {\it B}        & 11.83$\pm$0.03 & TIC v8.2\\
	    {\it V}        & 10.90$\pm$0.03 & TIC v8.2\\
	    {\it J}        &  9.16$\pm$0.03 & TIC v8.2\\
	    {\it H}        &  8.67$\pm$0.03 & TIC v8.2\\
	    {\it K$\rm_s$} &  8.58$\pm$0.02 & TIC v8.2\\
	    {\it W1}       &  8.52$\pm$0.02 & TIC v8.2\\
	    {\it W2}       &  8.61$\pm$0.02 & TIC v8.2\\
	    {\it W3}       &  8.52$\pm$0.03 & TIC v8.2\\
	    {\it W4}       &  8.24$\pm$0.03 & TIC v8.2\\
        Spectral class & K2V & P\&M$^{3}$\\
        $R(R_\odot$) & 0.78$\pm$0.01 & This work\\
        $M(M_\odot$) & 0.80$\pm$0.07 & This work\\
        $\log\,L(L_{\odot}$) & -0.489$\pm$0.008 & This work\\
        $A_{\rm V}$ & 0.015$\pm$0.009 & This work\\
        $Age(Gyr)$ & 9$\pm$6 & This work\\
        $T_{\rm eff}$ (K) & 5070$\pm$143 & This work\\
        log\,g & 4.6$\pm$0.3 & This work\\
        $\rm [Fe/H]$ & +0.20$\pm$0.08 & This work\\
        $\rm [\alpha/Fe]$ & 0.05$\pm$0.06 & This work\\
        $\xi$ (km\,s$^{-1}$) & 0.8$\pm$0.3 & This work\\
		\hline
\end{tabular}
\footnotesize{{\bf References:} 
[1] \citet{gaiaDR32023};  \citet{paegert2021}; [3] \citet{pecaut2013}}
\end{table}

%

\section{Observations}
\label{sec:observations}

\subsection{Photometry}

TOI-5076 was observed by {\it TESS} between August, 2021 and November, 2021. We used {\it TESS} full frame images (FFIs) to study this object. FFIs were calibrated by the {\it TESS} Science Processing Operations Center at NASA Ames Research Center \citep{jenkinsSPOC2016}. The {\it TESS} satellite imaged TOI-5076 during the fourth year of operation in three sectors: sector 42 (Camera 4, CCD 3), sector 43 (Camera 3, CCD 2), and sector 44 (Camera 1, CCD 4). Full frame images were acquired with a cadence of 10 min. In total, the satellite collected 10220 images of the target. The first image was taken on August 21, 2021 at 04:29 UT and the last image 
on November 5, 2021 at 22:27 UT. In total, four transits have been observed. Fig.~\ref{fig:tesscut} shows the target pixel file (TPF) for TOI-5076 relative to the cadence 175714 of sector 42. The image represents an area of $\sim$3.5 square arcmin around the target. The red dots denote sources from {\it Gaia} DR3 corrected for proper motion at the TPF epoch. Their dimension is scaled inversely and proportionally to their difference in apparent {\it Gaia} G-band magnitude with respect to the target. We used a modified version of \texttt{tpfplotter} \citep{aller2020} to generate this figure.

\subsection{High-angular-resolution imaging}

\begin{figure}
\includegraphics[width=\columnwidth]{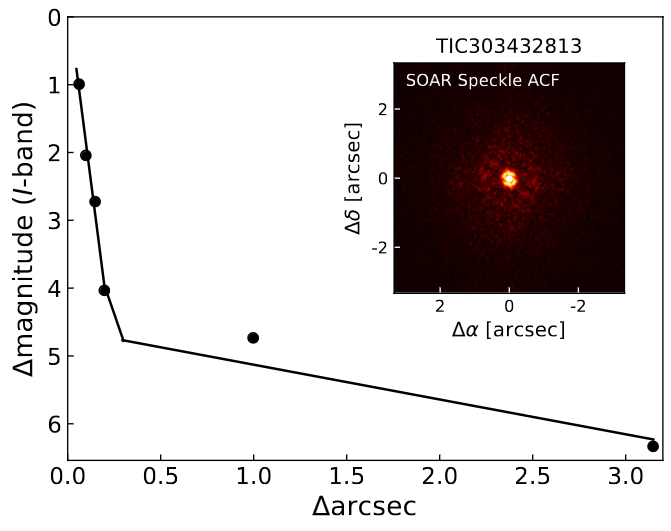}
\includegraphics[width=\columnwidth]{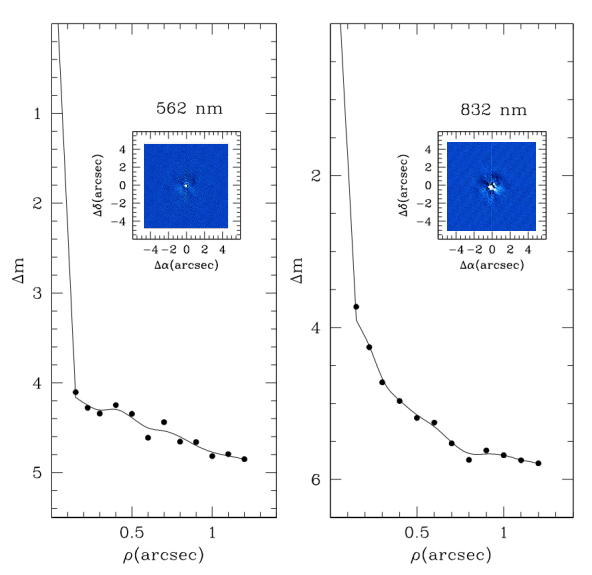}
\includegraphics[width=\columnwidth]{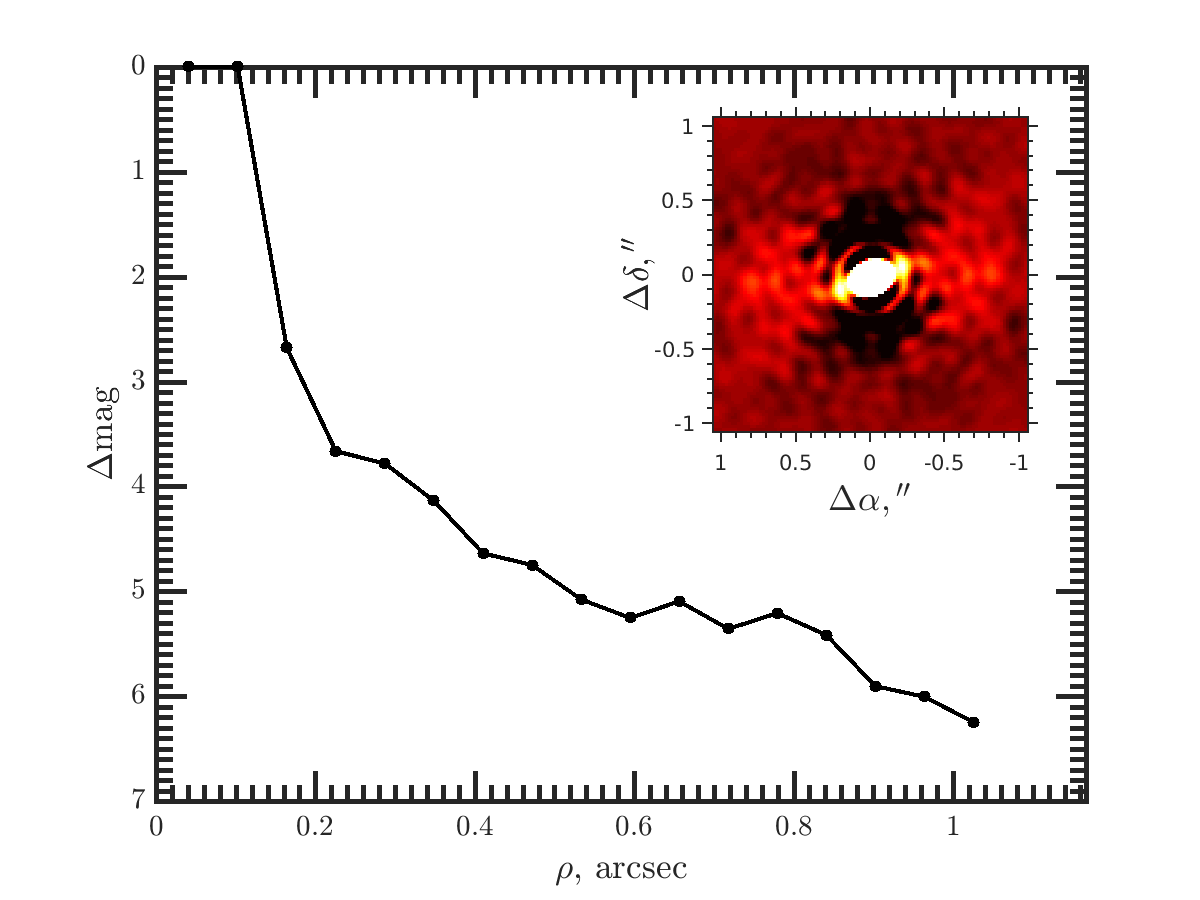}
\caption{
This figure presents high-angular-resolution imaging observations of the target.
Top: Speckle imaging obtained at 4.1-m Southern Astrophysical Research (SOAR) telescope. 
 Middle: Speckle imaging obtained at 3.5-m Wisconsin-Indiana-Yale-NOIRLab (WIYN) telescope.
 Bottom: Speckle imaging obtained at 2.5-m telescope at the Caucasian Observatory of the Sternberg Astronomical Institute (SAI). 
}
\label{fig:soar_sai}
\end{figure}

\subsubsection{SOAR} 

High-angular-resolution imaging is needed to search for nearby sources that can contaminate the {\it TESS} photometry, resulting in an underestimated planetary radius, or that may be the source of astrophysical false positives, such as background eclipsing binaries. We searched for stellar companions to TOI-5076 with speckle imaging on the 4.1-m Southern Astrophysical Research (SOAR) telescope \citep{tokovinin2018} on November 4, 2022 UT, observing in Cousins I-band, a similar visible band pass to {\it TESS}. This observation was sensitive to a 4.9-magnitude-fainter star at an angular distance of 1 arcsec from the target. More details of the observations within the SOAR {\it TESS} survey are available in \citet{ziegler2020}. The 5$\sigma$ detection sensitivity and speckle auto-correlation functions from the observations are shown in Fig.~\ref{fig:soar_sai} (top). No nearby stars were detected within 3\arcsec of TOI-5076 in the SOAR observations.

\subsubsection{WIYN} 

We used the NN-Explore Exoplanet Stellar Speckle Imager \citep[NESSI, ][]{scott2018} at the Wisconsin-Indiana-Yale-NOIRLab (WIYN) 3.5-m telescope to obtain high-resolution images of TOI-5076 on February 5, 2023 UT as part of the speckle imaging queue at WIYN.  NESSI is a dual-channel imager that takes simultaneous image sets in two cameras at red and blue wavelengths.  We used 40~nm wide filters with central wavelengths of 562~nm and 832~nm.  The target was centered in a 256$\times$256~pixel (4.6$\times$4.6~arcsec) subregion of the CCDs and observed in a series of 9000, 40~ms frames at a plate scale near 0.018$\rm\arcsec/pixel$ in each camera.  The TOI-5076 observations were followed by a 1000-frame observation of HD~23363 to serve as a point source calibrator.  Additionally, during the queue run, various binary stars with well-established astrometric properties were observed to calibrate the instrument's plate scales and orientations. The NESSI speckle data were reduced using a custom speckle data reduction pipeline described by \citet{howell2011}.  This pipeline produces high-level data products including reconstructed images of the field around the target and contrast curves measured from those images. We did not detect any stellar companions brighter than $\Delta$m = 4.85 and 5.79 at $\rho$ = 1.2$\rm\arcsec$ (and remaining constant out to 4.68 arcsec from the source) at 562~nm and 832~nm, respectively
(where $\rho$ is the separation from the source), as shown in Fig.~\ref{fig:soar_sai} (middle).

\subsubsection{SAI} 

TOI-5076 was observed on November 20, 2022 with the speckle polarimeter on the 2.5-m telescope at the Caucasian Observatory of Sternberg Astronomical Institute (SAI) of Lomonosov Moscow State University. The speckle polarimeter uses high--speed low--noise CMOS detector Hamamatsu ORCA--quest \citep{Strakhov2023}. The atmospheric dispersion compensator was active, which allowed us to use the $I_\mathrm{c}$ band. The respective angular resolution is 0.083\arcsec, while the long–exposure atmospheric seeing was 0.6\arcsec. We did not detect any stellar companions brighter than $\Delta I_\mathrm{c}=3.8$ and 6.2 at $\rho$ = 0.25\arcsec and 1.0\arcsec, respectively, where $\rho$ is the separation between the source and the potential companion (Fig.~\ref{fig:soar_sai}, bottom).


\subsection{Spectroscopy}

\subsubsection{TRES}

Two spectra were obtained using the Tillinghast Reflector Echelle Spectrograph \citep[TRES;][]{gaborthesis} on January 21,  2022 and January 31, 2022 as part of the TESS Follow-up Observing Program\footnote{\url{https://tess.mit.edu/followup}} Sub Group 1 \citep[TFOP;][]{tfop2019} to check for false positive scenarios and determine stellar parameters. 

During the first night, we obtained a spectrum with S/N=25.4 and a radial velocity of RV=71.145 km s$^{-1}$. During the second night, we obtained S/N=31.1 and RV=71.087 km s$^{-1}$. The two TRES observations yield a velocity offset of 58 m/s that was out of phase with the photometric ephemeris as reported in the ExoFOP observing notes
\footnote{\url{https://exofop.ipac.caltech.edu/tess/edit_obsnotes.php?id=303432813}}. TRES radial velocities were not used in the radial-velocity fit (Sect.~\ref{sec:planetary_parameters}).

TRES is mounted on the 1.5-m  telescope atop the Fred Lawrence Whipple Observatory (FLWO) in Arizona. It operates in the spectral range of 390 to 910 nm, with a spectral resolution of R=44,000.

\subsubsection{HARPS-N}

Spectroscopic observations were obtained with the HARPS-N spectrograph at the Telescopio Nazionale Galileo (TNG) at the Observatorio del Roque de los Muchachos (La Palma). HARPS-N is an echelle spectrograph covering the wavelength range from 383 to 693 nm, with a spectral resolution R=115000, which allows radial-velocity measurements with the highest accuracy currently available in the northern hemisphere. We acquired 44 measurements with an exposure time of 15 min, 20 min, or 30 min within the context of the GAPS program. The observations were performed between August 16, 2022 and March 16, 2023.

\begin{figure}
\includegraphics[width=\columnwidth]{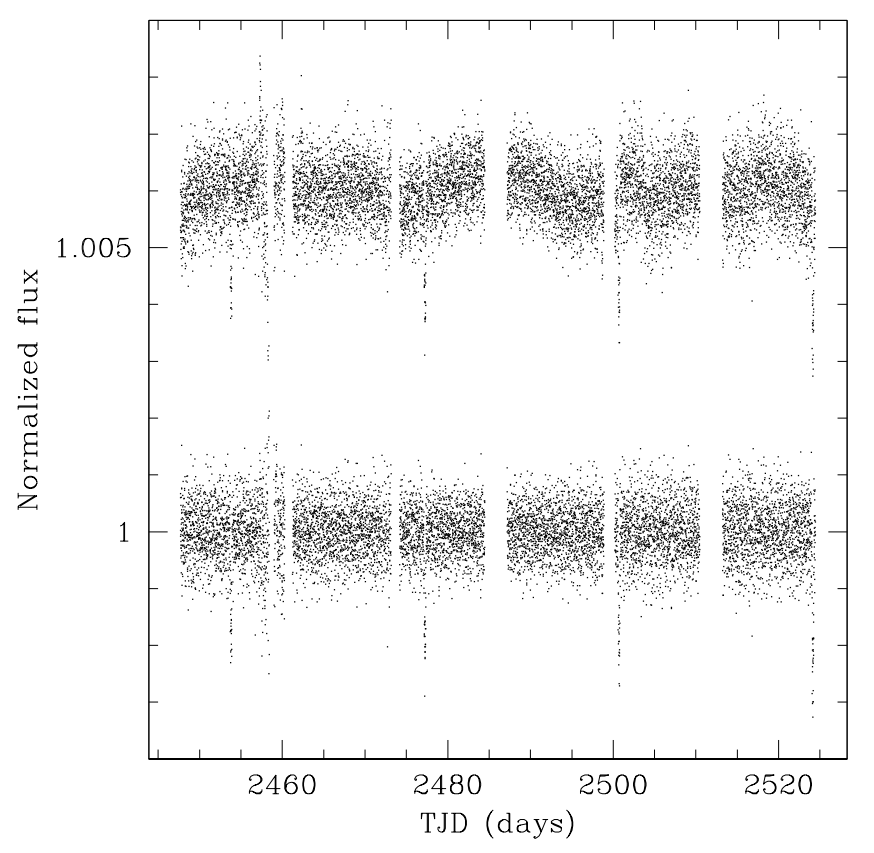}
\caption{
 This figure shows the light curve obtained from {\it TESS}
 during different processing steps.
 {\it Top:} {\it TESS} light curve corrected for systematics 
  with eigen light curves. {\it Bottom:} Final light curve 
  normalized by a spline fit on out-of-transit data. The 
  light curves are offset vertically by +0.006 for clarity.
}
\label{fig:TOI5076_full_lightcurve}
\end{figure}

\begin{figure}
\includegraphics[width=\columnwidth]{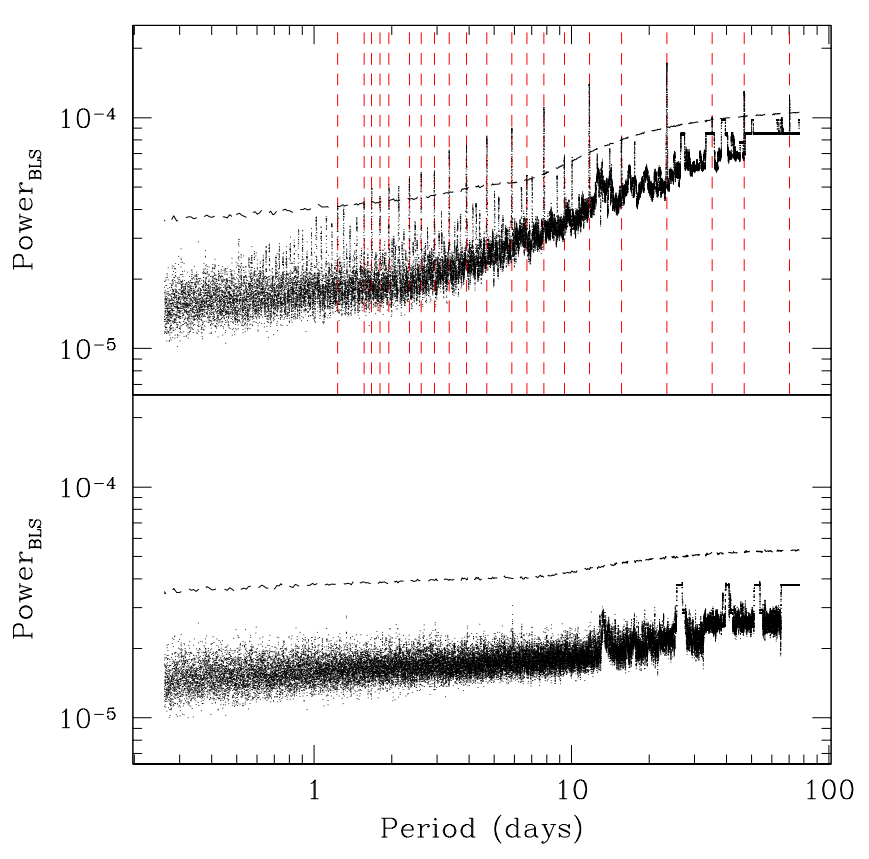}
\caption{
BLS periodogram of systematics-corrected and splined light curve (top) and of the same light curve with transits subtracted (bottom).
The dashed curves indicate the 1\% false-alarm-probability (FAP) threshold. The red, vertical, dashed lines in the top panel denote the periods where the BLS power surpasses the threshold.
}
\label{fig:powerBLS}
\end{figure}



\begin{table}
        \centering
        \caption{{\it TESS} photometric measurements of TOI-5076.}
        \label{tab:photometric_observations}
        \begin{tabular}{ccc} 
            \hline
         BJD$_{\textrm{TDB}}$ & Flux & $\rm\sigma_{Flux}$ \\
                \hline
         2459447.698153591    &       1.00002    &       0.00040 \\
         2459447.705098755    &       1.00067    &       0.00040 \\
         2459447.718989083    &       1.00060    &       0.00040 \\
         2459447.725934246    &       0.99992    &       0.00040 \\
          ...                 &       ...        &           ... \\
              \hline
        \end{tabular}
\end{table}

\begin{table*}
        \centering
        \caption{HARPS-N observations of TOI-5076.}
        \label{tab:spectroscopic_observations}
        \begin{tabular}{ccccccc} 
            \hline
        BJD & RV(km s$^{-1}$) & $\rm\sigma_{RV}$ (km s$^{-1}$) & BIS-SPAN(km s$^{-1}$) & CCF$\rm_{FWHM}$ (km s$^{-1}$) & $\log$ R$^{\prime}\rm_{HK}$ & $\sigma$($\log$ R$^{\prime}\rm_{HK}$)\\
                \hline
2459807.64277837 & 70.2756 & 0.0039 & -0.0158 & 6.9295 & -4.855 & 0.067 \\
2459807.72103835 & 70.2815 & 0.0029 & -0.0152 & 6.9056 & -4.999 & 0.058 \\
2459829.61427814 & 70.2787 & 0.0017 & -0.0285 & 6.9321 & -4.878 & 0.022 \\
2459830.71054652 & 70.2802 & 0.0021 & -0.0254 & 6.9243 & -4.879 & 0.030 \\
2459831.67995089 & 70.2739 & 0.0015 & -0.0268 & 6.9321 & -4.841 & 0.016 \\
2459832.64938936 & 70.2698 & 0.0025 & -0.0332 & 6.9405 & -4.824 & 0.032 \\
2459834.66033080 & 70.2714 & 0.0020 & -0.0287 & 6.9309 & -4.900 & 0.027 \\
2459835.59212502 & 70.2736 & 0.0017 & -0.0236 & 6.9248 & -4.850 & 0.020 \\
2459845.60891293 & 70.2902 & 0.0024 & -0.0225 & 6.9009 & -4.863 & 0.035 \\
2459856.67270407 & 70.2745 & 0.0026 & -0.0347 & 6.8923 & -4.923 & 0.047 \\
2459860.57438335 & 70.2745 & 0.0022 & -0.0338 & 6.8948 & -4.872 & 0.032 \\
2459868.64093323 & 70.2830 & 0.0022 & -0.0244 & 6.9189 & -4.900 & 0.033 \\
2459870.70237288 & 70.2836 & 0.0025 & -0.0202 & 6.9131 & -4.895 & 0.039 \\
2459871.57054841 & 70.2837 & 0.0018 & -0.0273 & 6.9199 & -4.840 & 0.021 \\
2459872.58566194 & 70.2873 & 0.0022 & -0.0306 & 6.8886 & -4.845 & 0.027 \\
2459873.59373655 & 70.2812 & 0.0017 & -0.0210 & 6.8964 & -4.880 & 0.023 \\
2459895.47214704 & 70.2842 & 0.0014 & -0.0318 & 6.9009 & -4.864 & 0.016 \\
2459899.44957031 & 70.2729 & 0.0046 & -0.0366 & 6.9125 & -4.987 & 0.103 \\
2459944.54390745 & 70.2802 & 0.0044 & -0.0245 & 6.9127 & -4.800 & 0.072 \\
2459950.55675591 & 70.2739 & 0.0046 & -0.0450 & 6.9405 & -4.758 & 0.067 \\
2459951.55191869 & 70.2764 & 0.0022 & -0.0351 & 6.9423 & -4.867 & 0.033 \\
2459952.44278344 & 70.2773 & 0.0021 & -0.0283 & 6.9174 & -4.799 & 0.026 \\
2459959.51669227 & 70.2827 & 0.0038 & -0.0165 & 6.9585 & -4.821 & 0.057 \\
2459960.46923467 & 70.2804 & 0.0055 & -0.0309 & 6.9570 & -4.679 & 0.069 \\
2459962.48603162 & 70.2854 & 0.0027 & -0.0189 & 6.9300 & -4.790 & 0.039 \\
2459963.48066282 & 70.2782 & 0.0033 & -0.0124 & 6.9469 & -4.850 & 0.054 \\
2459979.36983691 & 70.2807 & 0.0019 & -0.0238 & 6.9359 & -4.855 & 0.025 \\
2459989.40190871 & 70.2806 & 0.0032 & -0.0288 & 6.9187 & -4.812 & 0.047 \\
2459990.39532730 & 70.2786 & 0.0028 & -0.0200 & 6.9097 & -4.930 & 0.051 \\
2459994.39604873 & 70.2754 & 0.0025 & -0.0307 & 6.9375 & -4.886 & 0.039 \\
2459995.40764912 & 70.2798 & 0.0025 & -0.0224 & 6.9424 & -4.848 & 0.036 \\
2460005.34621237 & 70.2810 & 0.0017 & -0.0213 & 6.9289 & -4.822 & 0.021 \\
2460006.34977398 & 70.2824 & 0.0019 & -0.0243 & 6.9236 & -4.876 & 0.029 \\
2460007.36967721 & 70.2825 & 0.0014 & -0.0314 & 6.9273 & -4.836 & 0.015 \\
2460008.34435384 & 70.2832 & 0.0025 & -0.0266 & 6.9274 & -4.901 & 0.044 \\
2460009.34464210 & 70.2817 & 0.0013 & -0.0243 & 6.9294 & -4.846 & 0.015 \\
2460010.34901628 & 70.2799 & 0.0029 & -0.0400 & 6.9338 & -4.837 & 0.045 \\
2460011.33673759 & 70.2824 & 0.0015 & -0.0264 & 6.9242 & -4.835 & 0.017 \\
2460012.33230608 & 70.2817 & 0.0013 & -0.0306 & 6.9267 & -4.859 & 0.014 \\
2460013.33448350 & 70.2802 & 0.0030 & -0.0145 & 6.9236 & -4.809 & 0.046 \\
2460014.35733111 & 70.2704 & 0.0029 & -0.0253 & 6.9239 & -4.816 & 0.044 \\
2460018.36230230 & 70.2763 & 0.0018 & -0.0219 & 6.9348 & -4.804 & 0.022 \\
2460019.33861882 & 70.2769 & 0.0022 & -0.0320 & 6.9260 & -4.842 & 0.031 \\
2460020.34248018 & 70.2853 & 0.0029 & -0.0184 & 6.9121 & -4.809 & 0.046 \\
                \hline
        \end{tabular}
\end{table*}

\section{Data analysis}
\label{sec:data_analysis}

\subsection{Photometry}

The {\it TESS} images were analyzed with the \texttt{DIAmante} pipeline following the procedure described in \citet{montalto2020}. In brief, the {\it TESS} FFIs were analyzed with a difference imaging approach where a stacked reference image was subtracted from each image after convolving the reference by an optimal kernel. 
The photometry was extracted using a circular aperture of radius equal to 2 pixels. We chose this aperture after testing different aperture radii between 1 pix and 4 pix and selecting the one with the smallest scatter. The light curves from different sectors were merged together by accounting for sector-by-sector photometric zero-point variations; then, they were corrected for systematics on a sector-by-sector basis by using a best set of eigen light curves
extracted from the quietest and most highly correlated light curves in the same CCD where the target was observed \citep{montalto2020}.
To select the number of eigen light curves, we used the Akaike criterium, which indicated that three eigen light curves were appropriate. Finally, we normalized the light curve by a cubic spline function fitted on out-of-transit data with knots set every three hours. 
Table~\ref{tab:photometric_observations}, available at CDS,  presents the photometric measurements extracted by the pipeline. The first column indicates the time of observations at mid exposure (in BJD$\rm_{TDB}$), the second column the normalized flux (Flux), and the third column the error on the normalized flux ($\rm\sigma_{Flux}$).


In Fig.~\ref{fig:TOI5076_full_lightcurve}, we show the light curve corrected for systematics using eigen light curves (top) and the final splined light curve (bottom). We only analyzed the data that were not flagged by the pipeline \citep[see ][]{montalto2020} and that lie within a [+3$\sigma$, -5$\sigma$] interval between the mean of the normalized out-of-transit flux (9592 measurements). The transit analysis was further restricted to the measurements acquired between $\pm$10 hours from each transit event (721 measurements).
As can be seen in Fig. \ref{fig:tesscut}, three faint sources are included in the photometric aperture. Because the difference imaging approach subtracts all the constant sources and because the reference flux of the target is calibrated from the {\it TESS} \citep[TICv8.2,][]{paegert2022}
magnitude derived from {\it Gaia} (where the target is resolved from the contaminating sources), no dilution correction is needed.

\subsection{Spectroscopy}

The spectra of TRES were extracted following procedures outlined in 
\cite{buchhave2010}.
The spectroscopic data of HARPS-N were reduced by the HARPS-N Data Reduction 
Software \citep[DRS v3.7, ][]{lovis2007} using a G2 template mask.  
The first three columns of Table~\ref{tab:spectroscopic_observations} report the 
observation time at mid-exposure (in BJD),
the radial velocities (RV in km s$^{-1}$) and their 
errors ($\rm\sigma_{RV}$ in km s$^{-1}$) and are available at CDS.

\begin{figure}
\includegraphics[width=\columnwidth]{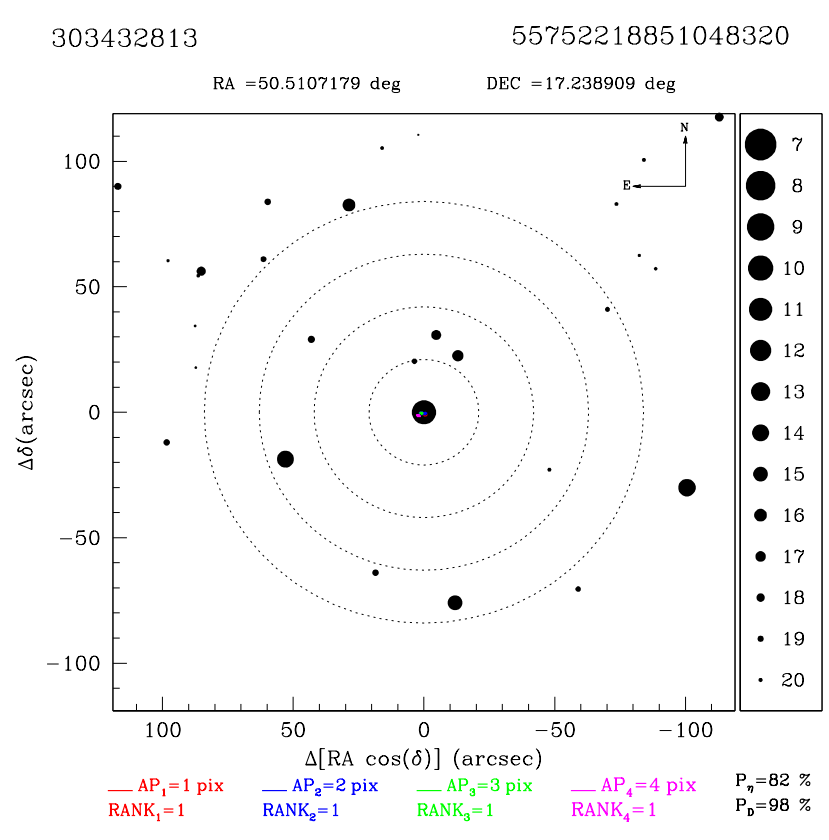}
\caption{
Centroid motion analysis of TOI-5076. The target (at the center) and the surrounding stars are represented by black dots with diameters that are proportional to their {\it Gaia} DR3 magnitudes, as indicated by the legend on the right. All positions were corrected for proper motions to report them to the epoch of cadence 175714 of Sector 42. The figure also reports the TIC identification number and the 
{\it Gaia} DR3 source ID (in the top left and top right, respectively), the distributions of the calculated centroid motion measurements in each aperture (central colored ellipses), the Mahlanobis distance ranks, and the centroid motion probabilities (P$_D$ and P$_{\eta}$; see text).
}
\label{fig: CENTROID_TOI5076}
\end{figure}


\section{Vetting}
\label{sec:vetting}

TOI-5076.01 was subjected to preliminary vetting to establish its plausibility as a planetary candidate. We used the vetting procedure reported in \citet{montalto2020} and \citet{montalto2023} in which a random-forest classifier was trained to distinguish between planetary candidates
and other false positives such as eclipsing binaries and short-term sinusoidal variables. The classifier received in input features extracted from the {\it TESS} light curve and the box-least-squares \citep[BLS,][]{kovacs2002}
periodogram. 
The BLS periodogram of the TOI-5076b systematics corrected and splined light curve (Fig,~\ref{fig:TOI5076_full_lightcurve}, bottom) is presented in Fig.~\ref{fig:powerBLS} (top). The highest peak is found at 23.444 days and it corresponds to a S/N = 35.9 for the transits folded with the estimated period. After modeling and subtracting the 23.444-day signal, the BLS periodogram did not show any other significant features associated with plausible transit events (Fig.~\ref{fig:powerBLS}, bottom).
The random-forest classifier calculated a probability (P$\rm_{RF}$) that the candidate is an exoplanet equal to P$\rm_{RF}$=99.6$\%$ corresponding to a false positive rate $<10^{-6}$. We also analyzed the centroid motion of the target during the transit events to establish if the transits were likely associated with the target star or to a star near the target. The centroid motion was analyzed in four concentric apertures between 1 pix and 4 pix, and two probabilities were derived: P$_D$, the probability that the
source of variability was associated with the target given the observed centroid motion measurements distributions and P$_{\eta}$, the reliability of the centroid motion results considering the local density of stars surrounding the target \citep{montalto2020,montalto2023}. For TOI-5076, we obtained
P$_D=98\%$ and P$_{\eta}=82\%$, which are both well above our adopted thresholds for acceptance (P$_D=50\%$ and P$_{\eta}=30\%$). Moreover, the algorithm also identified the star with the smallest Mahlanobis distance to the center of the distribution of centroid motion measurements of each aperture. For all tested apertures, we obtained that TOI-5076 was the star with the closest distance (RANK=1 for all apertures) as depicted in 
Fig.~\ref{fig: CENTROID_TOI5076}. From this analysis, we concluded that TOI-5076.01 is a statistically validated exoplanet, and hereafter we name it TOI-5076b.

\begin{figure}
\includegraphics[width=\columnwidth]{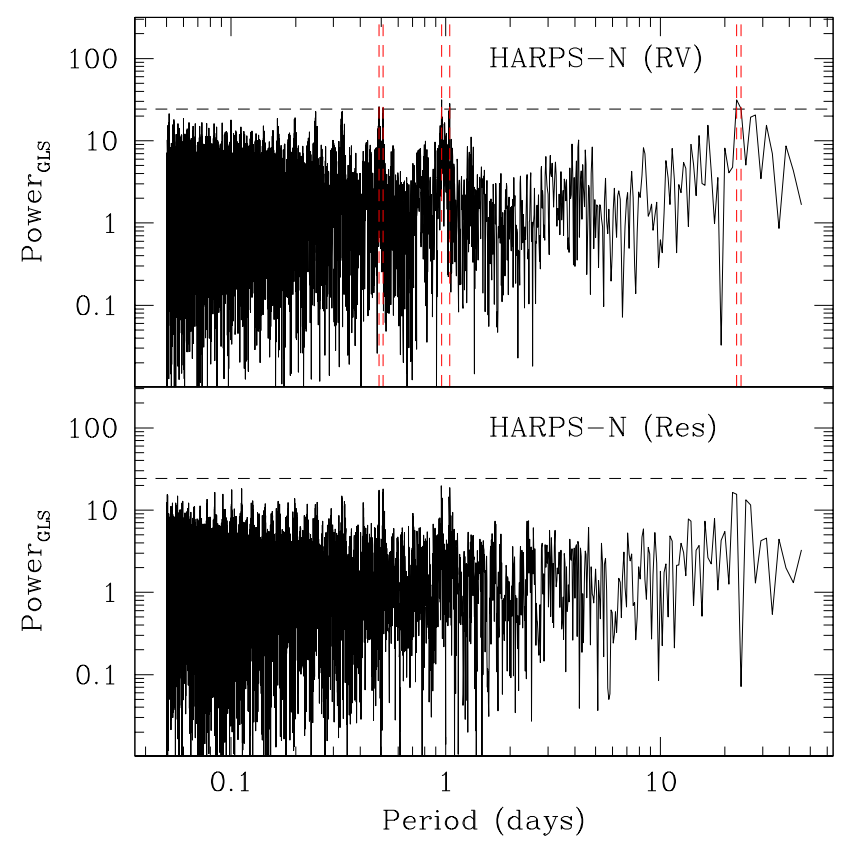}
\caption{
GLS periodogram of HARPS-N radial-velocity
measurements (top) and of the radial velocities measurements after subtraction of the best-fit radial-velocity model (bottom). The horizontal dashed line indicates the 1$\%$ FAP threshold.
The red, vertical, and dashed lines in the top panel denote the periods where the GLS power exceeds the FAP threshold.
}
\label{fig:powerGLS_HARPSN}
\end{figure}

In Fig.~\ref{fig:powerGLS_HARPSN} (top), we also present the generalized Lomb-Scargle \citep[GLS; ][]{zechmeister2009} periodogram of the HARPS-N radial-velocity measurements. The horizontal dashed line indicates the 1$\%$ false alarm probability threshold obtained from 1000 bootstrap samples of the residual radial-velocity measurements. The highest peak of the HARPS-N radial-velocity measurements presents a GLS peak at  22.740 days, which is close to the period inferred from the global modeling (Table~\ref{tab:system_parameters_A}) and above the 1$\%$ false-alarm threshold. We also note two peaks close to one day just above the 1$\%$ false-alarm-probability (FAP) threshold. However, after subtraction of the best-fit radial-velocity model (Sect.~\ref{sec:planetary_parameters}) the periodogram of the residual radial-velocity measurements (Fig.~\ref{fig:powerGLS_HARPSN}, bottom) does not present any significant peak above the 1$\%$ FAP threshold (as we already reported above for the analysis of the light curve residuals) indicating that no additional planets are detected in the data.

\section{Properties of the host star}
\label{sec:properties_of_the_host_star}

\subsection{Spectroscopic parameters}
\label{sec:spectroscopic_parameters}

We measured the star's effective temperature (T$\rm_{eff}$), surface gravity (log g),  iron abundance [Fe/H], and microturbulence velocity ($\xi$) using the equivalent width method. We used the software \texttt{StePar} \citep{tabernero2019}, which implements a grid of MARCS model atmospheres \citep{gustafsson2008} and the \texttt{MOOG} \citep{sneden1973} radiative transfer code to compute stellar atmospheric parameters by means of a downhill simplex minimization algorithm, which minimizes a quadratic form composed of the excitation and ionization equilibrium of Fe. Equivalent widths were measured with \texttt{ARES}~v2
\citep{sousa2015} from the  coadded spectrum obtained from the individual HARPS-N spectra used for the radial-velocity measurements.  The S/N of the coadded spectrum in the region between 5764~$\AA$ and 5766~$\AA$ measured using the ARES software was equal to S/N=280.
We used the FeI and FeII line list provided by the authors for the case of the Sun.
Using this approach, we obtained the parameters reported in Table~\ref{tab:stellar_properties}. 

We also used the Stellar Parameter Classification tool \citep[SPC][]{buchhave2012} to derive stellar parameters using the TRES recon spectra and the average results from the two spectra (Teff = 4947 K $\pm$ 50 K, log g = 4.54 $\pm$ 0.10, [m/H] = 0.23 $\pm$ 0.08, and vsini = 3 km s$^{-1} \pm$ 1  km s$^{-1}$) agree with the results reported in Table~\ref{tab:stellar_properties}.

We also note that, for TOI-5076, the Str\"{o}mgren photometry is available in \citet{hauck1998} which provide the following: (b-y)=0.576, m1=0.589, c1=0.260. 
We corrected for reddening the photometry using E(B-V)=0.005 obtained from our estimated
extinction A$\rm_V=0.015$ (Sect.~\ref{sec:stellar_parameters}) adopting a standard reddening law
and the relations $\rm x_0 = x - E(B-V) * k_{x}$ where x=(b-y),m1,c1 and k$\rm_x$=0.822,-0.319,0.176 are deduced from \citet{arnadottir2010}. We then obtained (b-y)$\rm_0$=0.572, m1$\rm_0$=0.591, c1$\rm_0$=0.259.
Adopting the calibration of \citet{ramirez2005}, we obtained an estimated photometric metallicity equal to
[Fe/H]=+0.35. The estimated uncertainty on the metallicity derived from this photometric calibration is 0.2 dex \citep{ramirez2005}. Overall, this result agrees with the spectroscopic metallicity and confirms that the target is a metal-rich star.

\subsection{$\alpha$-elements}

Using the same radiative transfer code and model atmospheres as done for the determination of the stellar atmospheric parameters and iron abundance, and considering the line list by \cite{Biazzo2022}, we measured the elemental abundances of the four 
$\alpha$-elements Mg, Si, Ca, and Ti. Using these elements and the definition [$\alpha$/Fe]=$\frac{1}{4}$([Mg/Fe]+[Si/Fe]+[Ca/Fe]+[Ti/Fe]), 
we obtain [$\alpha$/Fe]=0.05$\pm$0.06\footnote{As claimed by \cite{adibekyan2011}, for [Fe/H]$>$0 the [Ca/Fe] trend for dwarf stars 
in the Galactic disk differs from that of Mg, Si, and Ti. Also excluding this element for the [$\alpha$/Fe] computation, we obtain 
[$\alpha$/Fe]=(0.06$\pm$0.06), which is consistent with the value obtained considering all four elements.} for TOI-5076.


\subsection{Empirical spectral library}
\label{sec:empirical_spectral_library}

\begin{figure}
\includegraphics[width=\columnwidth]{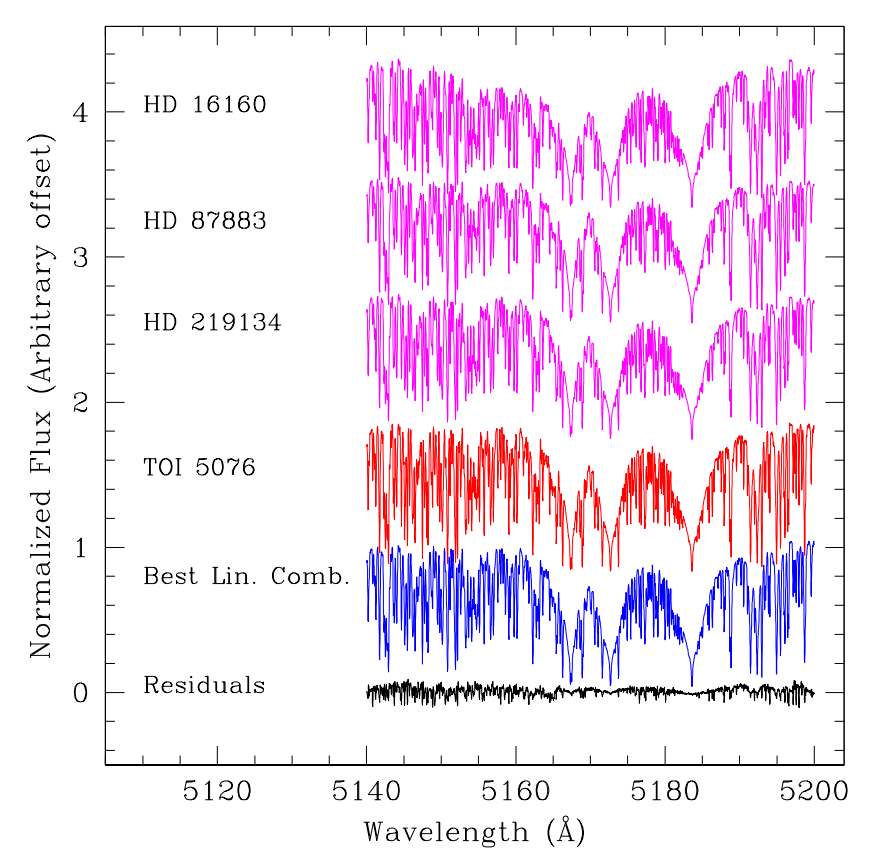}
\caption{
Comparison of target spectrum in Mgb triplet region with empirical
spectra in the \citet{yee2017} library. The target's spectrum is depicted in red. The three spectra of the library most highly correlated with the target's spectrum are shown in magenta, while the best-fit
linear combination of them is depicted in blue. The residuals between the target's spectrum and the best fit are represented in black. 
}
\label{fig:TOI5076_EmpiricalLibrary.png}
\end{figure}

We also compared the spectroscopic parameters 
we derived in the previous section with the ones obtained using the spectra of the empirical library
of \citet{yee2017}. This library includes 404 stars observed with Keck/HIRES by the California Planet Search. We used the software \texttt{SpecMatch-Emp}
\citep{yee2017} to perform the comparison between our stacked spectrum of TOI-5076 and the library spectra. In Fig.~\ref{fig:TOI5076_EmpiricalLibrary.png}, we represent the three empirical library spectra\footnote{The spectra of stars HD~16160, HD~87883, and HD~219134.} most highly correlated with the target spectrum
in the region of the Mgb triplet in magenta. We indicate the target spectrum in red and the best-fit linear combination of the three reference spectra reported above in blue. Finally, at the bottom we show the difference between the target's spectrum and the linearly combined reference spectra in black. In this case we obtained T$\rm_{eff}$=(4728$\pm$110) K, 
log$\,$g=(4.5$\pm$0.1) dex, and [Fe/H]=(0.10$\pm$0.09) dex, which are all consistent within 2$\sigma$ with the spectroscopic parameters previously derived and adopted. 

\begin{table*}
        \centering
        \caption{System parameters relative to TOI-5076.}
        \label{tab:system_parameters_A}
        \begin{tabular}{llccc}
                \hline
                Parameter & Symbol & Value & Priors & Units \\
                \hline
                {\it Fitted parameters} & & & & \\
                Transit Epoch (TJD) & $T_0$ & 2430.366$\pm$0.005 & $\mathcal{U}$(2430.30,2430.40) & days \\
        Orbital Period & $P$ & 23.445$\pm$0.001 & $\mathcal{U}$(23.40,23.50) & days \\
        r1 parameter \citep{espinoza2018a} & $\rm r1$ & 0.44$\pm$0.09 & $\mathcal{U}$(0,1) & \\
        r2 parameter \citep{espinoza2018a} & $\rm r2$ & 0.037$\pm$0.001 & $\mathcal{U}$(0,1)& \\
                Stellar reflex velocity & $K$ & 4.1$\pm$0.6 & $\mathcal{U}$(-100; 100) & m s$^{-1}$ \\
            Center-of-mass velocity & $\gamma$ & 70.2792$\pm$0.0004 & $\mathcal{U}$(-100; 100) & km s$^{-1}$\\
            Stellar density & $\rho_{\star}$ & 1.4$\pm$0.2& $\mathcal{N}$(2.377; 0.6), $0.25<\rho_{\star}<5.00$  & g cm$^{-3}$\\
            Radial velocity jitter (HARPS-N) & $\sigma_{\textrm{HARPS-N}}$ & 1.9$\pm$0.5 & $\ln\sigma_{\textrm{HARPS-N}}\in\mathcal{U}$(0.001; 1000) & m s$^{-1}$ \\
        Offset relative flux for \textrm{TESS} & \textrm{$\mu_{TESS}$}& 0.00002$\pm$0.00003 & $\mathcal{N}$(0, 0.1) & \\
            Jitter error (TESS) & $\sigma_{\textrm{TESS}}$ & 288$\pm$28 & $\ln\sigma_{\textrm{TESS}}\in\mathcal{U}$(0, 1000) & ppm \\
            Parameter related to linear limb darkening (TESS) & $q_{1,\textrm{TESS}}$ &  0.7$\pm$0.2 & $\mathcal{U}$(0; 1) \\
            Parameter related to quadratic limb darkening (TESS) & $q_{2,\textrm{TESS}}$ & 0.7$\pm$0.2 & $\mathcal{U}$(0; 1) \\
            \hline
            {\it Derived parameters} & & & & \\
        Planet-to-star radius ratio & $p=\frac{R_p}{R_{\star}}$ & 0.037$\pm$0.001 & - & - \\
                Impact parameter & $b$ & 0.2$\pm$0.1 & - & - \\
            Orbital inclination & $i$ & 89.7$\pm$0.2 & - & $^{\circ}$\\
            Scaled semi-major axis of the orbit  & $\frac{a}{R_{\star}}$ & 34$\pm$1 & - & - \\    
            Planet mass &  $m_p$ &  16$\pm$2 & - & M$\rm_{\oplus}$\\
            Planet radius & $r_p$  & 3.2$\pm$0.1 & - & R$\rm_{\oplus}$\\
            Planet surface gravity & $log\,g_{p}$ & 3.2$\pm$0.1 & - & - \\
            Planet density & $\rho_p$ & 2.8$\pm$0.5 & - & g cm$\rm ^{-3}$\\
            Planet equil. temp. (A=0) & $T_{eq}$ & 615$\pm$20 & - & K \\
        Total duration of transit & $T\rm_{41}$ & 5.2$\pm$0.2 & - & hr \\
            Linear limb darkening (TESS) & $u_{1,\textrm{TESS}}$ & 1.1$\pm$0.3 & - & - \\
            Quadratic limb darkening (TESS) & $u_{2,\textrm{TESS}}$ & -0.3$\pm$0.3 & - & - \\
                \hline
        \end{tabular}
  {\footnotesize NOTE -- The eccentricity was set to zero, and the argument of periastron was set to 90 degrees. An eccentric orbit solution is discussed in Sect.~\ref{sec:planetary_parameters}.} \\
\end{table*}

\subsection{Stellar parameters}
\label{sec:stellar_parameters}

We derived stellar parameters using 
 {\it Gaia} DR3 \citep{riello2021}, 2MASS \citep{skrutskie2006, cohen2003}, and ALLWISE (W1, W2, W3, and W4) photometry \citep{wright2010, jarrett2011} together with the spectroscopic constraints obtained in Sect.~\ref{sec:properties_of_the_host_star}. The effective temperature, the gravity, and the metallicity were constrained using a Gaussian prior and adopting the values reported in Table~\ref{tab:stellar_properties}.
 We imposed a uniform prior on the distance,  [$d$-3$\times\,\sigma_{d}$; $d$+3$\times\,\sigma_{d}$], where 
 $d=82.78$ pc was the distance  obtained from the simple inversion of the parallax, and $\sigma_d=0.07$ pc was the semi-difference between the upper and lower distance estimates obtained by subtracting and adding the parallax standard error from and to the {\it Gaia} DR3 parallax value. 
 We also imposed a uniform prior on the interstellar extinction $A\rm_{550}$ (the monochromatic extinction at 550 nm). We first calculated the expected value of the monochromatic extinction ($A\rm_{550}$) for our target
 using the extinction map of \citet{lallement2022}. 
 At the position of TOI-5076, we obtained $A\rm_{550}=$(0.015$\pm$0.005). Then we considered the values between [0; $A\rm_{550}$+3$\,\times\,\sigma_{A\rm_{550}}$]=[0; 0.030] as a plausible interval for the optical extinction.

 The broadband photometry was obtained using the Padova library of stellar isochrones \citep[PARSEC,][]{bressan2012}. We generated a set of isochrones with an age ($\tau$) comprised between $\tau$= [10$^7$; 10$^{10}$] yr and a logarithmic step size equal to log(Age) = 0.5. We varied the metallicity of the isochrone set between [M/H] = [-2.0; 0.3] in steps of size equal to 0.05 dex. For each value of the effective temperature, gravity, and metallicity generated by the algorithm, we identified the stellar model with the closest stellar parameters in our isochrone set. 
 A small perturbation was applied to the effective temperature and the gravity to estimate the gradient of each quantity of interest (photometry and stellar parameters) with respect to these two parameters in a close surrounding of the identified model grid point, and a first-order Taylor expansion was applied to estimate any physical quantity of interest to the exact value of the simulated gravity and effective temperature.
 We then obtained the corresponding broadband absolute photometry and applied the distance modulus and extinction. 
  To convert the monochromatic extinction ($A\rm_{550}$) to the extinction in any other photometric band ($A\rm_X$), we calculated the ratio ($A\rm_X$/$A\rm_{550}$) using a set of MARCS \citep{gustafsson2008} stellar spectra of dwarf stars for a broad range of effective temperatures and extinctions and interpolated the results for a star of 
  T$\rm_{eff}$=5070 K and a $A\rm_{550}$=0.015.
 We also calculated the parallax (from the simulated distance value). 
 
 We then compared these simulated values of the broadband photometry and of the parallax with the observed ones. The log-likelihood function we adopted to evaluate the model performance was equal to 
 $\ln\mathcal{L}=-\rm\frac{1}{2}\sum_{i=1}^{i=Nobs} (\frac{o_{\rm i}-s_{\rm i}}{\sigma_{o_{\rm i}}})^2$, where $o_{\rm i}$ and $\sigma_{o_{\rm i}}$ 
 denote the observed magnitudes (and uncertainties) in different photometric bands and the measured parallax (and uncertainty), while the $s_{\rm i}$ are the corresponding simulated quantities.
 For any simulated model, we also registered the value of the stellar mass, stellar radius, luminosity, and age.
To incorporate a systematic error component in our analysis, we also performed the fit using the set of BaSTI isochrones \citep{hidalgo2018} adding the differences
of the best-fit parameters' values obtained using, respectively, the Padova
and BaSTI models in quadrature to the parameters' uncertainties obtained using the Padova models.

\begin{figure}
\includegraphics[width=0.9\columnwidth]{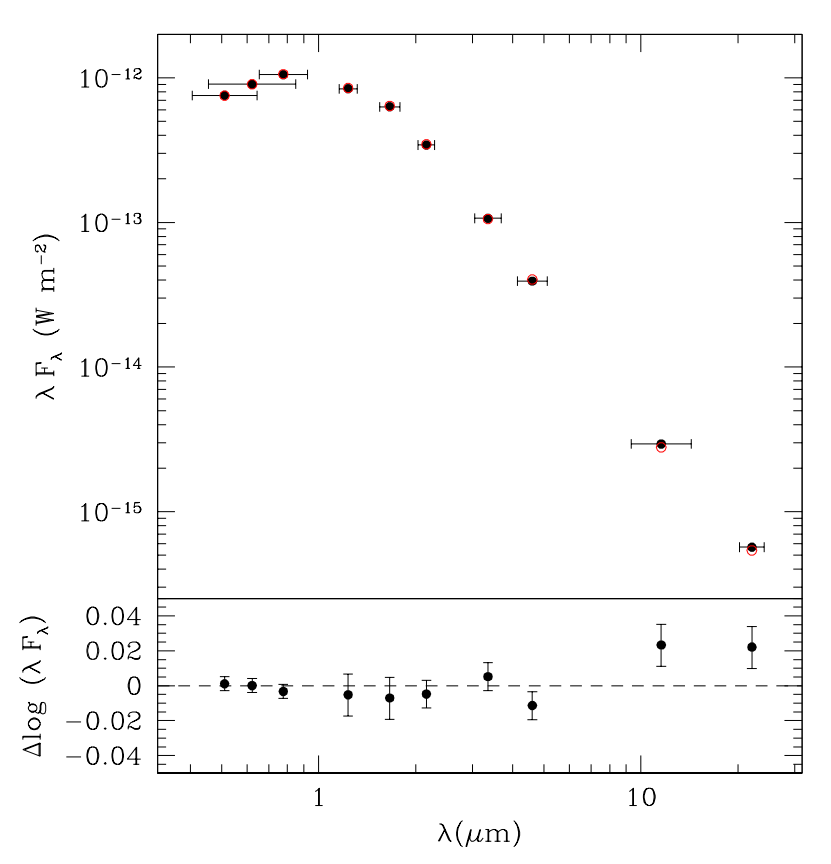}
\caption{
This figure presents the spectral energy distribution of the target obtained from broadband optical and near-infrared photometry.
 {\it Top:} Optical and near-infrared broadband photometric measurements of the target star (black circles) and best-fit model (red open circles).
 {\it Bottom:} Difference between logarithm of measurements and model's estimates.
}
\label{fig:SED}
\end{figure}

 The posterior distributions of the parameters were obtained using the nested sampling method implemented in the \texttt{MultiNest} package \citep{feroz2008,buchner2014,feroz2009,feroz2019}. We used 250 live points. The result of the fit is shown in Fig.~\ref{fig:SED}. The reduced chi-square of the fit ($\rm\chi_r$) is equal to $\rm\chi_r$=1.1\footnote{We added an error of 0.01 mag in quadrature to the \textit{Gaia} broadband photometric errors reported in Table~\ref{tab:stellar_properties}.}. The best-fit stellar mass and radius are M$_{\star}$=(0.80$\pm$0.07) M$_{\odot}$ and R$_{\star}$=(0.78$\pm$0.01) R$\rm_{\odot}$, respectively.
 We also obtained a distance (d) equal to d=(82.8$\pm$0.1) pc, an extinction equal to A$\rm_V$=(0.015$\pm$0.009), and an age equal to $\tau$=(9$\pm$6) Gyr.
 The results of our analysis are reported in Table~\ref{tab:stellar_properties}. 

\begin{figure}
\includegraphics[width=\columnwidth]{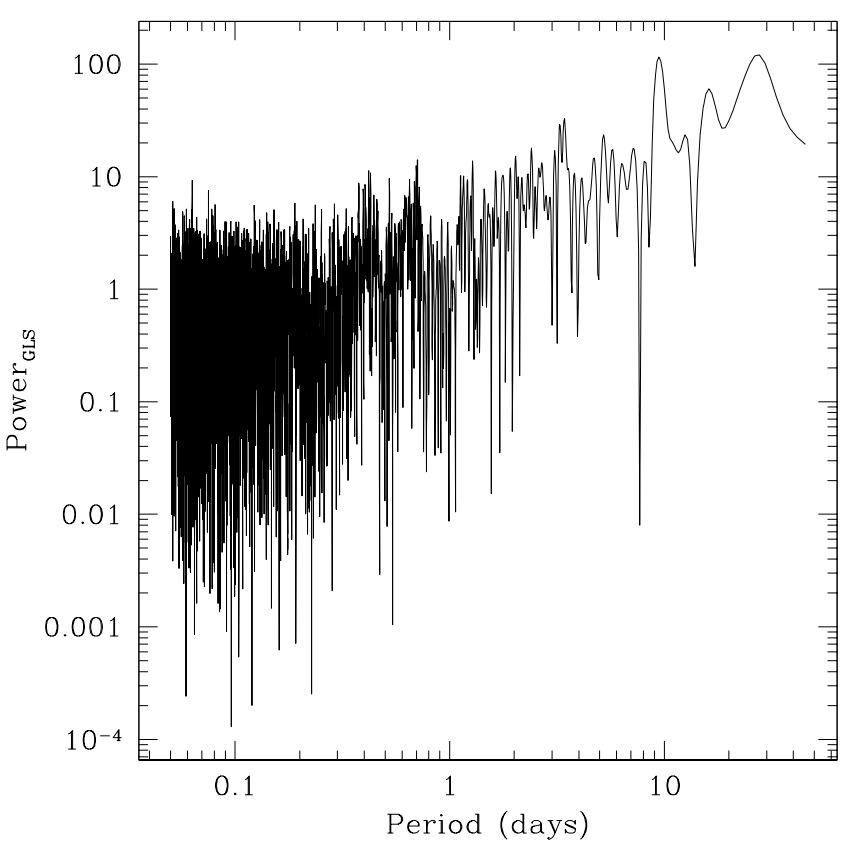}
\caption{
Generalized Lomb-Scargle periodogram of systematics corrected out-of-transit data obtained by TESS.
}
\label{fig:oot}
\end{figure}

Using the empirical spectral library described in Sect.~\ref{sec:empirical_spectral_library},
we also derived the stellar parameters obtaining
R$_{\star}$=(0.8$\pm$0.1) R$_{\odot}$, M$_{\star}$=(0.77$\pm$0.08) M$_{\odot}$,  and log Age=(12.5$\pm$5) Gyr, which are all consistent with our previous estimate. Moreover, the template spectra (Fig.~\ref{fig:TOI5076_EmpiricalLibrary.png}) used to match the spectrum of TOI-5076 all have v$\,$sin$\,$i=1 km$\,$s$^{-1}$. 
Such a value is at the limit of the HARPS-N resolution. 
The generalized Lomb-Scargle periodogram \citep[GLS;][]{zechmeister2009} of the systematics-corrected out-of-transit data obtained by {\it TESS} (Fig.~\ref{fig:oot}) has the highest peak at $\sim$27.79 days. This is also close to the period of 33.4 days inferred from the activity index-rotation period relationship (see Sect.~\ref{sec:chromospheric_activity}) of \citet{astudillo2017}.
However, since the duration of a \textit{TESS} sector is about 27 days, the observed modulation should be considered with care.

An alternative method to determine the stellar parameters is described in \citet{montalto2021} and employed for the construction of the all-sky PLATO input catalog (as PIC1.1). In this case, we obtained
T$\rm_{eff}$=(4867$\pm$206) K,
R$_{\star}$=(0.81$\pm$0.09) R$_{\odot}$ and M$_{\star}$=(0.79$\pm$0.07) M$_{\odot}$.
These estimates are all compatible with the parameters obtained by the Bayesian approach described in the first part of this section, which was finally adopted in our analysis along with the stellar parameters we derived.

\subsection{Binarity}
\label{sec:binarity}

The host star forms a common proper motion pair with the close-by companion 2MASS J03220140+1714465 (\textit{Gaia} DR3 55752974765291264) as reported in \citet{gould2004}. \textit{Gaia} DR3 supports such an association giving for the companion star $\mu_{\alpha}=(168.20\pm0.08)$ mas yr$^{-1}$,
$\mu_{\delta}=(-175.41\pm0.07)$ mas yr$^{-1}$ and a parallax 
equal to $\pi=(11.99\pm0.07)$ which are consistent with the proper motions and parallax of TOI-5076 reported in Table~\ref{tab:stellar_properties}.
Using the \textit{Gaia} DR3 color and the method described in \citet{montalto2021}, we deduced that the companion star has an effective temperature of T$\rm_{eff}=(3115\pm190)$ K and is a M4.5V star.
The companion is located at a parallactic angle equal to PA=330.4$^{\circ}$ at a projected separation of 26.4 arcsec from TOI-5076, which corresponds to 2178 au.

\subsection{Galactic velocities}
\label{sec:galactic_velocities}

We used the kinematic approach of \citet{bensby2003} to calculate
the Galactic space velocities of TOI-5076 with respect to the local standard of rest (LSR) obtaining U$\rm_{LSR}$=-64.19 km s$^{-1}$, V$\rm_{LSR}$=-74.69  km s$^{-1}$, W$\rm_{LSR}$=-39.10  km s$^{-1}$ and to estimate the probability that the host star belongs to the thin disk ($\rm P_{thin}$), the thick disk ($\rm P_{thick}$), and the 
halo ($\rm P_{halo}$). For the probabilities, we obtained $\rm P_{thin}=5.1$\%, $\rm P_{thick}=94.8$\%, and $\rm P_{halo}=0.1$\%. We therefore conclude that TOI-5076 is likely a thick-disk star from a kinematic point of view, and therefore it should be an old star with an age on the order of (10$\pm$5) Gyr \citep[][]{bensby2003}.

\subsection{Stellar activity}
\label{sec:stellar_activity}

\subsubsection{Chromospheric activity}
\label{sec:chromospheric_activity}

In Fig.~\ref{fig:rv_bis_logR_fwhm} (top), we present the chromospheric activity index log R$^{\prime}\rm_{HK}$ as a function of the radial-velocity measurements. To calculate this index, we used the YABI software deployed at IA2 Data Center\footnote{ \url{https://ia2.inaf.it.}}. The Pearson correlation coefficient between the log R$^{\prime}\rm_{HK}$ index and the radial velocities is equal to 0.06 (p-value=0.6997), indicating a negligible correlation between these two quantities. The average log R$^{\prime}\rm_{HK}$ is  <log R$^{\prime}\rm_{HK}$>=(-4.85$\pm$0.05).  According to the classification proposed by \citet{henry1996}, TOI-5076 is inactive.
In Fig.~\ref{fig:resrv_bis_logR_fwhm} (top), we present the chromospheric activity index log R$^{\prime}\rm_{HK}$ as a function of the radial-velocity residuals
(calculated after subtraction of the best-fit radial-velocity model, Sect.~\ref{sec:planetary_parameters}).
The Pearson correlation coefficient between the log R$^{\prime}\rm_{HK}$ index and the radial velocities is equal to 0.001 (p-value=0.9938), indicating a negligible correlation between these two quantities.

\subsubsection{FWHM of the cross-correlation function}
\label{sec:fwhm}

In Fig.~\ref{fig:rv_bis_logR_fwhm} (middle), we present the FWHM of the cross-correlation function (CCF) as a function of the radial-velocity measurements.  
To calculate this index, we used the HARPS-N pipeline.
The Pearson correlation coefficient between the FWHM index and the radial velocities is equal to -0.2546 (p-value=0.0954), indicating a negligible correlation between these two quantities. 
In Fig.~\ref{fig:resrv_bis_logR_fwhm} (middle), we present the FWHM of the
CCF as a function of the radial-velocity residuals. 
The Pearson correlation coefficient between the FWHM index and the radial-velocity residuals is equal to -0.2498 (p-value=0.1019), indicating a negligible correlation between these two quantities. 

\subsubsection{Bisector span}
\label{sec:bisector_span}

\begin{figure}
\includegraphics[width=\columnwidth]{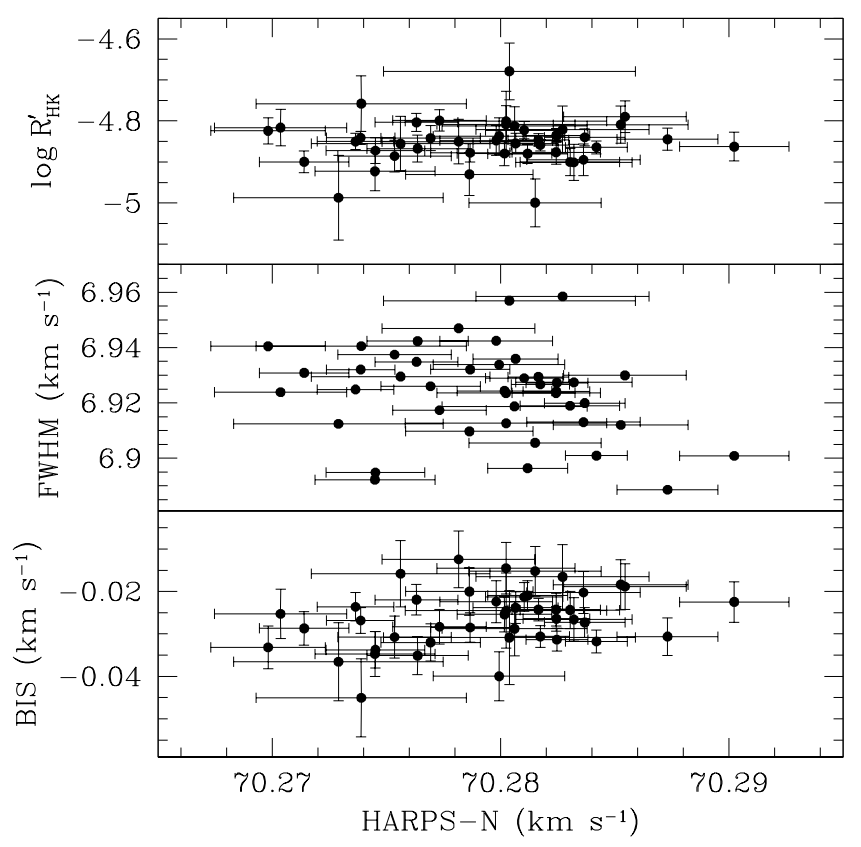}
\caption{
This figures shows the correlation between several activity indicators and the radial velocities.
\emph{Top: } Chromospheric activity index log~R~$^{\prime}_{HK}$ of TOI-5076 versus HARPS-N radial velocities. 
\emph{Middle: } FWHM of CCF versus radial velocities. \emph{Bottom: }  Bisector span versus radial velocities.
}
\label{fig:rv_bis_logR_fwhm}
\end{figure}

In Fig.~\ref{fig:rv_bis_logR_fwhm} (bottom), we report the bisector span \citep{queloz2001} versus the radial-velocity measurements. Such a quantity was calculated by the HARPS-N pipeline, and it is a measure of the line asymmetry, which may suggest the presence of issues related to activity and/or multiplicity.  The error on the bisector has been assumed equal to twice the error on the radial velocities. We measured only a weak, not significant, positive correlation between the radial velocities and the bisector span (r$\rm_{pearson}$=0.3393, p-value=0.02423). In Fig.~\ref{fig:powerGLS_BIS},
we show the periodogram of the bisector-span measurements and the false alarm probability threshold (FAP=1$\%$) calculated from 1000 bootstrap resamples
of the bisector-span measurements. No significant peak is observed.

\begin{figure}
\includegraphics[width=\columnwidth]{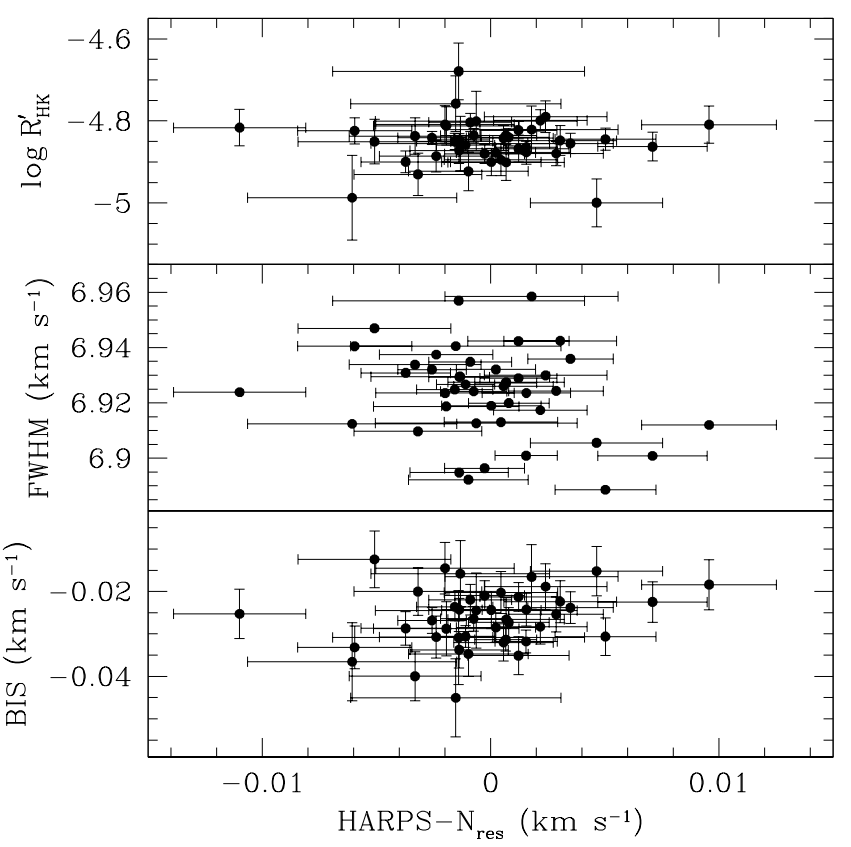}
\caption{
This figures shows the correlation between several activity indicators and the radial-velocity residuals.
\emph{Top: } Chromospheric activity index log R$ ^{\prime}_{HK}$ of TOI-5076 versus radial-velocity residuals
(obtained after subtraction of the best-fit radial-velocity model). 
\emph{Middle: } FWHM of CCF versus radial-velocity residuals.
\emph{Bottom: } Bisector span versus radial-velocity residuals.
}
\label{fig:resrv_bis_logR_fwhm}
\end{figure}

\begin{figure}
\includegraphics[width=\columnwidth]{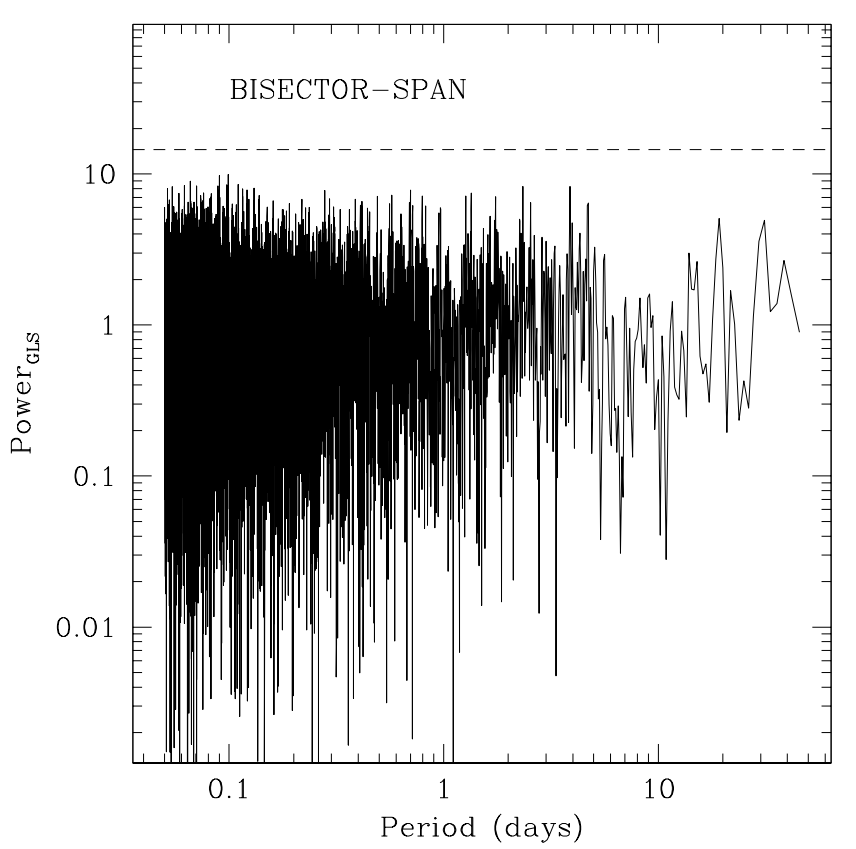}
\caption{
Periodogram of bisector-span measurements and FAP threshold
(FAP=1$\%$).
}
\label{fig:powerGLS_BIS}
\end{figure}

In Fig.~\ref{fig:resrv_bis_logR_fwhm} (bottom), we report the bisector span versus the radial-velocity residuals.
The Pearson correlation coefficient between the bisector span index and the radial-velocity residuals is equal to 0.2403 (p-value=0.1161), indicating a negligible correlation between these two quantities. On the basis of these results, we conclude that stellar activity has a negligible
impact on the observed radial velocities. 


\section{Planetary parameters}
\label{sec:planetary_parameters}

Planetary parameters were obtained by performing a simultaneous fit
of both spectroscopic and photometric data with the \texttt{juliet} software \citep{espinoza2019}. For the transiting planet, we assumed a Keplerian orbit, fixing the eccentricity to $e=0$ and the argument of periastron to $\omega=90^{\circ}$.
To model the planet-to-star radius ratio $p$ and the impact parameter $b,$ we adopted the parametrization of \citet{espinoza2018a}. 
We sampled limb-darkening  coefficients with the method described in \citet{kipping2013}.  We adopted uninformative priors for all parameters. 
Jitter parameters were adopted both for HARPS-N and for \textit{TESS} data to absorb underestimated white-noise errors and unaccounted-for sources of red noise. In total, we fit 12 parameters reported in the top section of Table~\ref{tab:system_parameters_A}.
The posterior distributions of the parameters were obtained using \texttt{MultiNest} via  \texttt{PyMultiNest}\citep{buchner2014}. We adopted 250 live points. 

We also fit a Keplerian orbit with nonzero eccentricity by fitting the additional parameters $\sqrt{e}\cos{\omega}$ and $\sqrt{e}\sin{\omega}$\footnote{We adopted a uniform prior $\mathcal{U}$(-0.89; 0.89) for both of them. 
}. The result
implied a log evidence of 
$\ln$ $Z_e=2975.82\pm0.03$ against a log evidence for the case of circular orbit of $\ln$ $Z_c=2976.3\pm0.1$. Consequently, the circular orbit was preferred with an odds ratio equal to $\exp(\Delta\ln Z)=\exp(\ln Z_c - \ln Z_e)=$1.6$\pm$0.2. The eccentric orbit implied an eccentricity $e=(0.20\pm0.09)$. The circular orbit solution is preferred in  our analysis because of its formally better log evidence with respect to the eccentric orbit solution
and the consistency of the eccentric orbit with a circular orbit solution within 2.2-$\sigma$.
We note that an eccentric orbit would imply a value for the stellar density equal to $\rho_{\star}=(2.1\pm0.5)$~g~cm$^{-3}$ and consistent within 0.5-$\sigma$ with the stellar density obtained from the estimation of the stellar mass and radius, which is equal to $\rho_{\star}=(2.4\pm0.4)$ g cm$^{-3}$  (Sect.~\ref{sec:stellar_parameters}). From the circular orbit solution, we obtained $\rho_{\star}=(1.4\pm0.2)$ g cm$^{-3}$ (Table~\ref{tab:system_parameters_A}), which is consistent within 2.2-$\sigma$ with the stellar density obtained from the estimation of the stellar mass and radius. 

\begin{figure}
\includegraphics[width=\columnwidth]{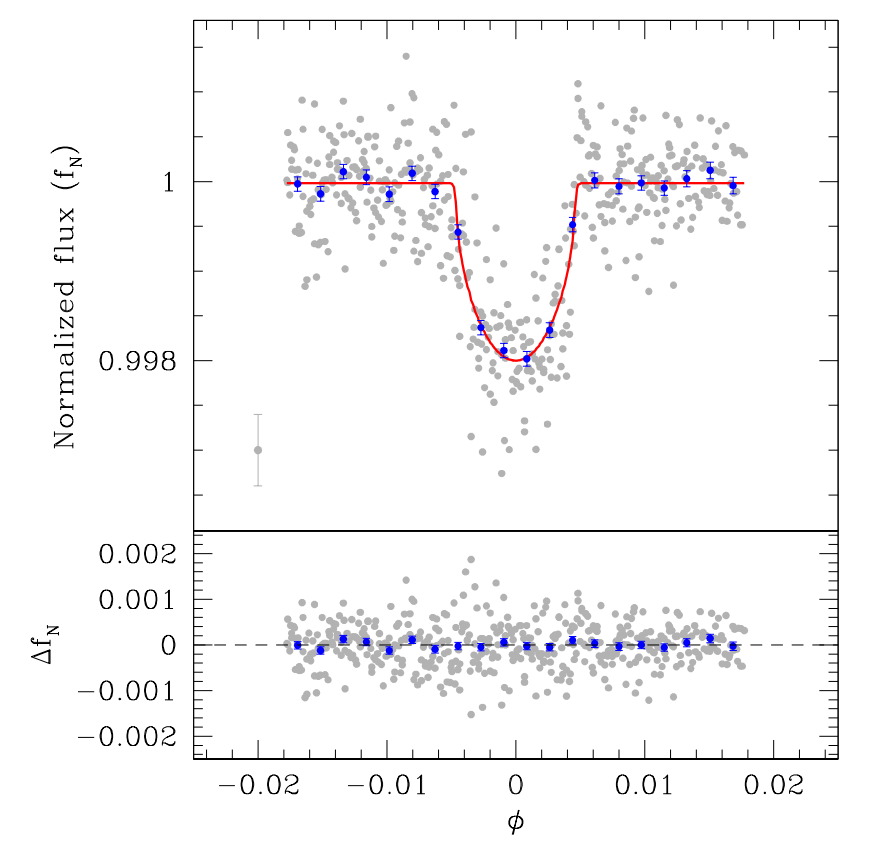}
\caption{
This figure shows the folded photometric measurements, the best-fit photometric model and the residuals of the fit.
\emph{Top:} {\it TESS} light curve of TOI-5076 (gray points) folded with best-fit ephemerides. The best-fit transit model is denoted by the red curve. The point in the bottom left depicts the average uncertainty of the photometric measurements. Blue points represent measurements rebinned in 20 phase bins between the minimum and maximum phases. \emph{Bottom:} Residuals of fit. 
}
\label{fig:TOI5076_lc_folded}
\end{figure}

\begin{figure}
\includegraphics[width=\columnwidth]{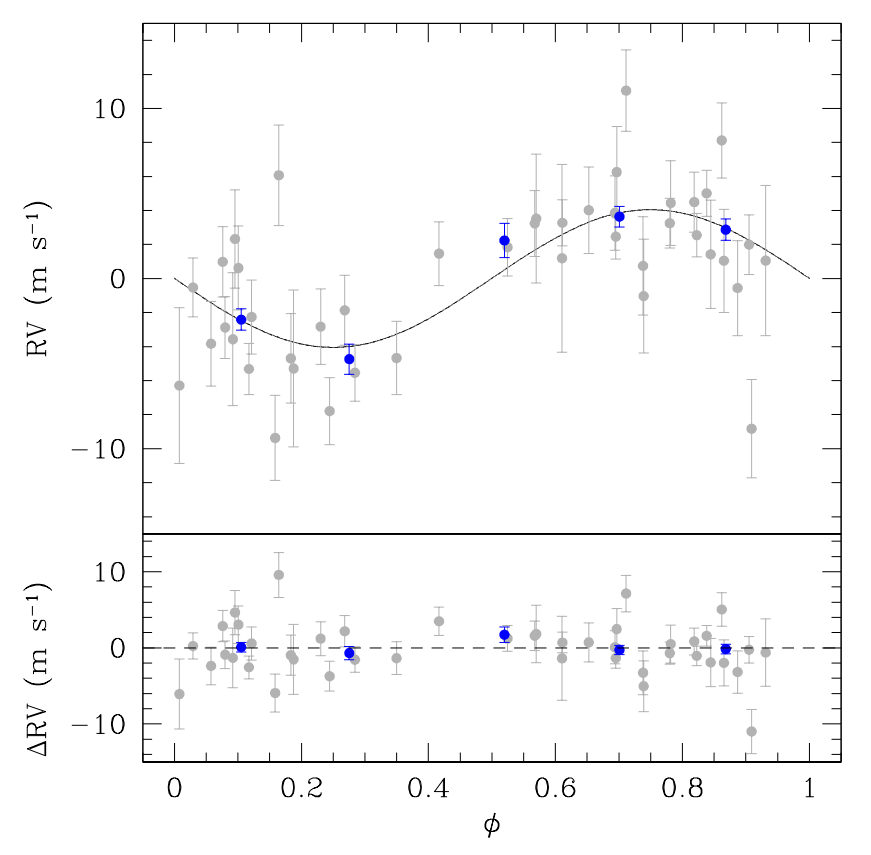}
\caption{
This figure shows the folded radial-velocity measurements, the best-fit radial-velocity model and the residuals of the fit. \emph{Top:} HARPS-N radial-velocity measurements (gray points) folded with the best-fit orbital period and time of transit. The continuous line denotes the best-fit radial-velocity model. Blue points represent measurements rebinned in five phase bins. \emph{Bottom:} Residuals of fit.
}
\label{fig:HARPS_RV}
\end{figure}

We obtained that the transiting body is a sub-Neptune planet with a mass m$\rm_p=$(16$\pm$2) M$\rm_{\oplus}$ and a radius r$\rm_p=$(3.2$\pm$0.1) R$\rm_{\oplus,}$ yielding a density $\rho_p$=(2.8$\pm$0.5) g cm$^{-3}$. The best-fit photometric and radial-velocity models are shown 
in Fig.~\ref{fig:TOI5076_lc_folded} and Fig.\ref{fig:HARPS_RV},
together with the photometric and radial-velocity measurements
folded with the best-fit period.

\begin{figure}
\includegraphics[width=\columnwidth]{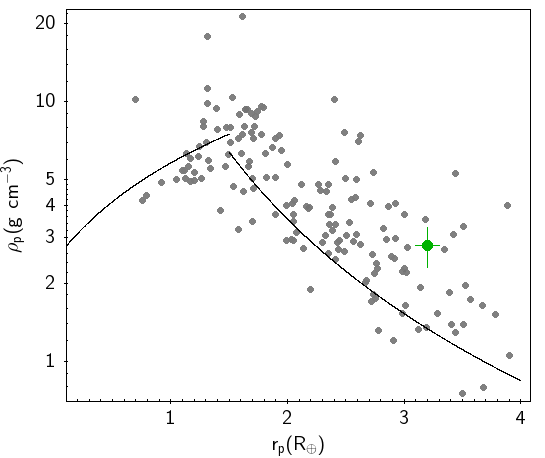}
\caption{
Planetary density versus planetary radius for  sample of planets with R$<$4 R$\rm_{\oplus}$ described in the text. The black continuous lines denote the relations of \citet{weiss2014}. The position of TOI-5076b in this diagram is indicated by the green dot.
}
\label{fig:RhopRp}
\end{figure}

\begin{figure}
\includegraphics[width=\columnwidth]{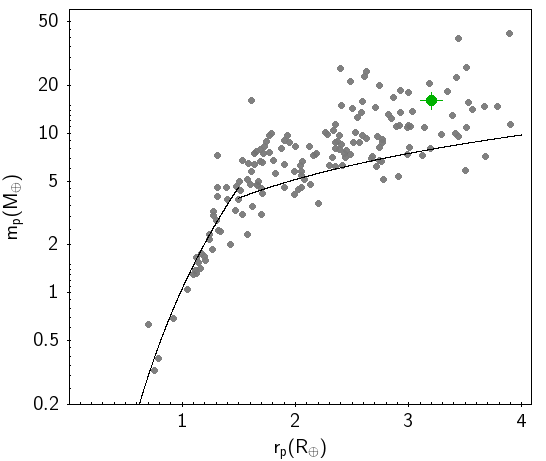}
\caption{
Planetary mass versus planetary radius for sample of planets with R$<$4 R$\rm_{\oplus}$ described in the text. The black continuous lines denote the relations of \citet{weiss2014}. The position of TOI-5076b in this diagram is indicated by the green dot.
}
\label{fig:mass_radius}
\end{figure}

\section{Discussion}
\label{sec:discussion}

The lack of correlation between the radial velocities and several activity indicators (Sect.~\ref{sec:stellar_activity}) supports the Keplerian origin of the RV variations
and the planetary nature of the transiting object.
TOI-5076b is a sub-Neptune planet with a mass 10$\%$ smaller and a radius
16$\%$ smaller than Neptune. This yields a density about 1.5 times that of Neptune.

\citet{weiss2014} studied the masses and radii of a sample of 65 exoplanets with radii of r$\rm_p$<4 R$_{\oplus}$ and orbital periods of less than 100 days. They found that
on average planets up to 1.5 R$_{\oplus}$ increase in density with increasing radius, indicating that their composition is mainly rocky. Planets with a radius of R~$>$1.5 R$_{\oplus}$ have average densities rapidly decreasing with increasing radius, suggesting that these planets have a large fraction of volatiles by volume overlying a rocky core. Moreover, they present a large compositional diversity of their atmospheres or rocky cores of different masses or compositions. 

In Fig.~\ref{fig:RhopRp}, we present the average density versus planetary radius diagram of a sample of exoplanets, which we retrieved from the NASA exoplanet archive\footnote{\url{https://exoplanetarchive.ipac.caltech.edu/index.html}}. We only considered confirmed planets discovered either by the radial-velocity technique or the transit method and with radii smaller than 4 R$_{\oplus}$. We imposed that both their radii and masses be determined  with a precision level better than 20$\%$. 
When multiple entries were present for the same planet, we used the default set of parameters adopted in the catalog by setting \texttt{default\_set}=1.
In total, the sample is composed of 165 exoplanets. The continuous lines in Fig.~\ref{fig:RhopRp} represent the relationships defined in \citet{weiss2014}, where for planets with r$\rm_p$<1.5 R$_{\oplus,}$

\begin{equation}
\rm \rho_p=2.43+3.39\,\Big(\frac{r_p}{R_{\oplus}}\Big)\,\rm\,g\,cm^{-3}
;\end{equation}

\noindent
while, for planets with 1.5 R$\rm_{\oplus} \leq r_p<4$ R$_{\oplus,}$

\begin{equation}
\rm \frac{m_p}{M_{\oplus}}=2.69+\,\Big(\frac{r_p}{R_{\oplus}}\Big)^{0.93}.
\label{eq:weiss2014}
\end{equation}

\noindent
The position of TOI-5076b in this diagram is indicated by the green dot and it appears consistent with the decreasing branch of the planetary density versus planetary radius relationship
defined by Eq~\ref{eq:weiss2014}.

In Fig.~\ref{fig:mass_radius}, we present the planetary mass versus planetary radius diagram for the same sample of exoplanets of Fig.~\ref{fig:RhopRp}. Also in this case, the black continuous lines indicate the relationships of \citet{weiss2014}, and the position of TOI-5076b in this diagram is indicated by the green dot. In this case, Eq.~2 would predict m$\rm_p$=7.9$\rm\,M_{\oplus}$
for a radius equal to 3.2 R$_{\oplus,}$
as does the one of TOI-5076b. The measured mass is instead m$\rm_p=(16\pm2)\,M_{\oplus,}$ which corresponds to 4.05$\rm\sigma_{m_p}$ above the predicted value. We may argue that, since the host star is metal-rich
(Tab.~\ref{tab:stellar_properties}), there could have been favorable conditions for the accretion of a 
massive rocky core or for polluting the atmosphere of TOI-5076b with heavy elements. However, by looking at the sample of planets represented in Fig.~\ref{fig:mass_radius}, TOI-5076b is not an extremely massive sub-Neptune planet since several other planets with similar radii have masses beyond that of TOI-5076b.

We note that the presence of inflated planets could affect the radii and density distributions in Fig.~\ref{fig:RhopRp} and Fig.~\ref{fig:mass_radius}.
There are 17 planets with radii larger than our target in the sample we considered. Only four of them have an orbital period shorter than three days, and therefore they could be expected to have high equilibrium temperatures and potentially inflated atmospheres (the rest having orbital periods longer than six days). These are TOI-849b \citep{armstrong2020}, TOI-2196b \citep{persson2022}, TOI-132b \citep{diaz2020}, and Kepler-94b \citep{marcy2014}. All these planets have the fact that they lie in the so-called Neptune desert in common \citep{mazeh2016}, which origin is still debated in the literature. These objects could possess extended atmospheres (as in the case of Kepler-94b) or could be the remnant cores of gas-giant planets (e.g., TOI-849b), or maybe the results of other complex evolutionary patterns.
In any case, we note that the different evolutionary histories of these planets could affect their radii and therefore could slightly influence the \citet{weiss2014} relations and change the relative position of those models with respect to our target position. However, this does not alter our conclusion  that TOI-5076b is not an extremely massive sub-Neptune planet since the mass of these planets is not significantly affected by the inflation or envelope mass-loss mechanism.

\begin{figure}
\includegraphics[width=\columnwidth]{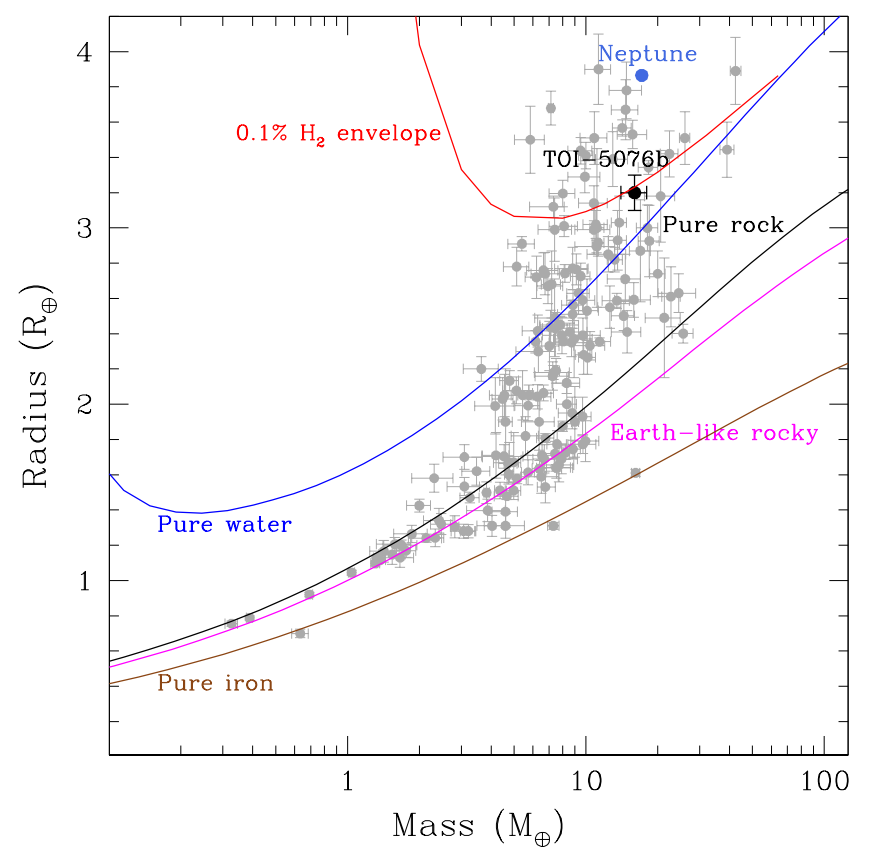}
\caption{
Radii and masses of exoplanets with r$\rm_p<$ 4 R$_{\oplus}$ discussed in the text (gray points). The position of TOI-5076b in this diagram is shown by the black dot. We also represent the position of Neptune (blue dot). The colored and continuous lines denote mass-radius models for planets with the same composition. In particular: pure iron (100\% Fe, brown curve); Earth-like rocky (32.5\% Fe+67.5\% MgSiO$_3$, magenta); pure rock (100\% MgSiO$_3$, black curve);
pure water (100 \% H$_2$O, blue curve);  
0.1\% H$_2$ envelope (49.95\% Earth-like rocky core + 49.95\% H$_2$O layer + 0.1\% H$_2$ envelope by mass), assuming 1 milli-bar surface pressure level and isothermal atmosphere at 700K (red curve).
}
\label{fig:RadiusVSMass}
\end{figure}

In Fig.~\ref{fig:RadiusVSMass}, we show the radius versus mass diagram for the sample of 165 planets discussed above (gray points) with a superimposed representative set of mass - radius models for planets with identical composition.\footnote{All models were taken from \url{https://lweb.cfa.harvard.edu/~lzeng/planetmodels.html}}
The position of TOI-5076b in this diagram seems more compatible with models with 
an H$_2$ envelope, such as the model
with a 49.95\% Earth-like rocky core and a 49.95\% H$_2$O layer and 0.1\% H$_2$ envelope by mass represented by the red curve rather than models that have a pure composition of only one element.

\begin{figure}
\includegraphics[width=\columnwidth]{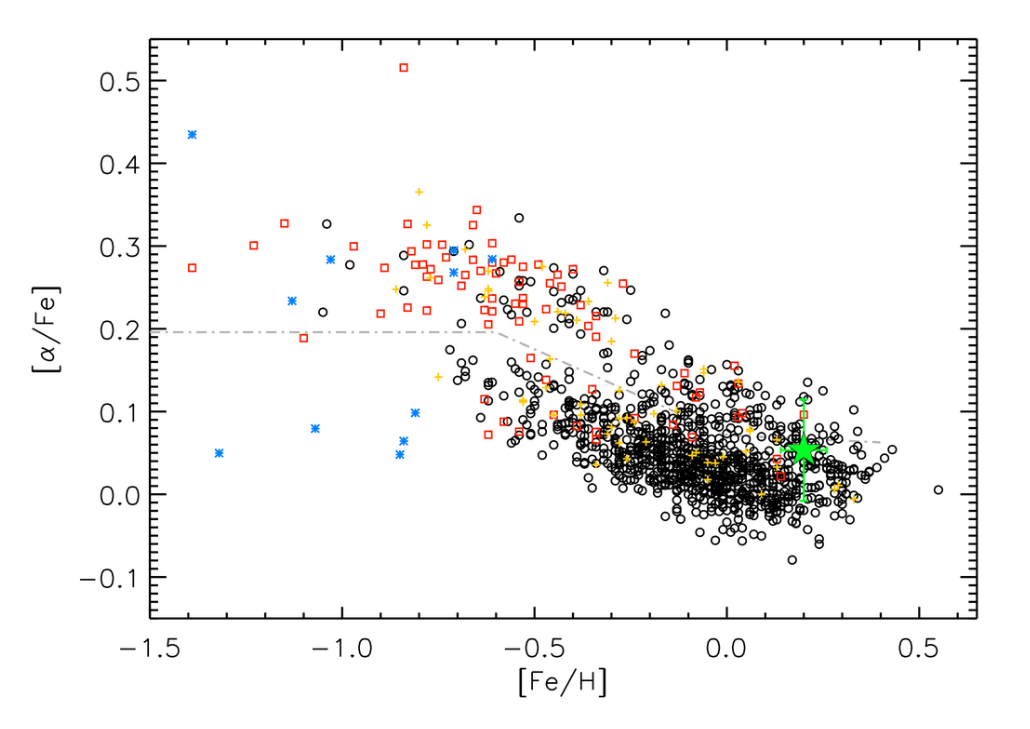}
\caption{
[alpha/Fe] vs [Fe/H] for TOI-5076 (star  symbol). Thin-disk, thick-disk, and halo field stars as in \citep{adibekyan2012b} are shown with circles, squares, and asterisks, respectively. The crosses refer to their thin-thick and thick-halo transition stars. The dot-dashed line separates the stars with high-and low-alpha content as derived by the same authors.\\
}
\label{fig:alpha_fe}
\end{figure}

In Sect.~\ref{sec:binarity}, we reported that the host star belongs to a binary system where the companion lies at a projected separation of 2178 au from the target.
Such a wide binary likely does not significantly affect the architecture of a planetary system around one of its components (a so-called S-type configuration)
as shown by statistical studies of planetary systems around wide (a$>$100-300 au) and close binaries, apart from the average eccentricity, which appears larger for planets in binary systems (both close and wide) than for planets around single stars \citep[e.g., ][]{su2021}. 
Some hints that the orbit of TOI-5076b could be eccentric 
come from the fact that the circular orbit solution implies a rather low stellar density in comparison to the theoretical stellar density expected from stellar models, whereas the stellar density obtained assuming an eccentric orbit is higher and more consistent with the stellar model's estimation, as we note in Sect.~\ref{sec:planetary_parameters}. 
Using the [$\alpha$/Fe] versus [Fe/H] positions as chemical indicators of stellar population (Fig.~\ref{fig:alpha_fe}), TOI-5076 appears to belong to the region populated 
by $\alpha$-enhanced and metal-rich stellar populations \citep{adibekyan2011, Bensbyetal2014} and is close to the transition region between low- and high-$\alpha-$ content stars (dot-dashed line in Fig.~\ref{fig:alpha_fe}).

\begin{figure}
\includegraphics[width=\columnwidth]{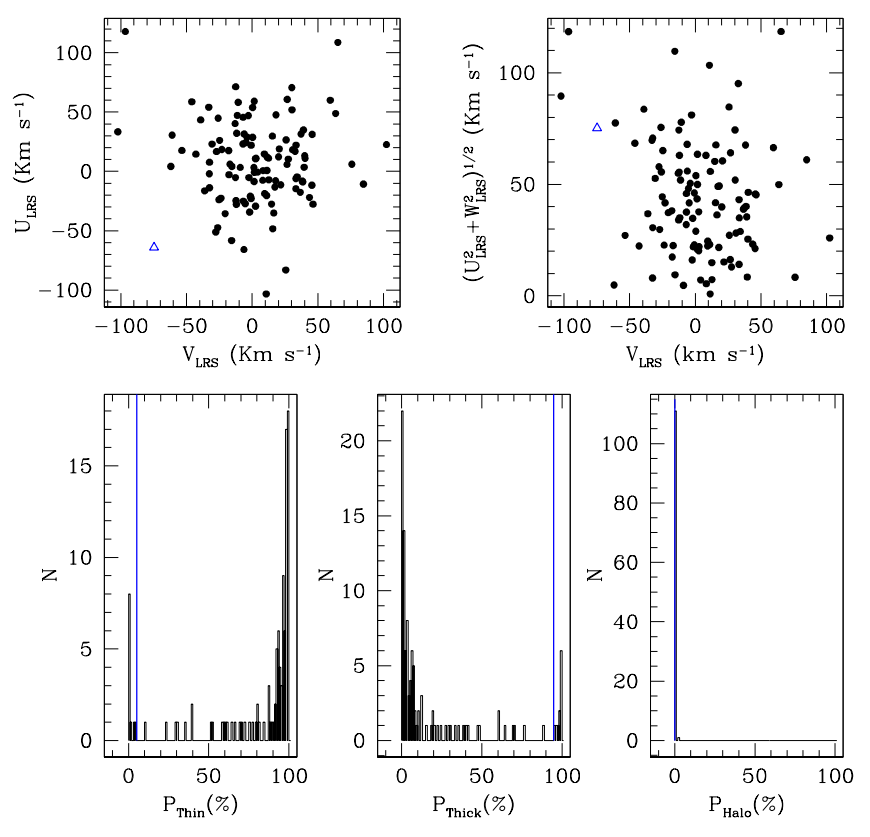}
\caption{
Boettlinger (top left) and Toomre (top right) diagrams for the sample of planet host stars (black dots) described in the text. The position of TOI-5076b is marked by the open blue triangle. In the bottom row are depicted the probability distributions to belong to the thin disk (bottom left), the thick disk (bottom middle) and the halo (bottom right). The vertical blue lines denote the probabilities for TOI-5076b.
}
\label{fig:Diagrams}
\end{figure}

This result, together with the values of the 
Galactic-space velocities and the high thick-to-thin disk probability reported in Sect.~\ref{sec:galactic_velocities} seem to be consistent with the possible belonging of TOI-5076 
to the old population of thin-to-thick-disk transition stars with relatively hot kinematics (e.g., see Fig. 20 in \citealt{Bensbyetal2014}).
This is also apparent in Fig.~\ref{fig:Diagrams}, where we present the Boettlinger and Toomre diagrams (top left and right, respectively) of the sample of planet host stars hosting at least one exoplanet with r$\rm_p<4$R$_{\oplus}$
(and for which a precision level in the estimation of planetary masses and radii better than 20$\%$ was achieved). This sample was retrieved from the NASA exoplanet archive, and it was matched with the ${\it Gaia}$ DR3 catalog from which we extracted the positions, proper motions, and radial velocities of the host stars. The distances were taken from the NASA exoplanet archive. In total, the sample amounts to 112 stars.\footnote{A few stars for which no systemic radial velocity was found neither in the {\it Gaia} or NASA catalogs nor in the literature were eliminated.} With these quantities, we calculated the Galactic velocities \citep{bensby2003} reported in the diagrams of Fig.~\ref{fig:Diagrams}. The position of TOI-5076b (blue open triangle) appears peripheral in these diagrams, indicating that the velocity of TOI-5076b is relatively high with respect to the velocities of the sample of stars studied here. In the bottom row of Fig.~\ref{fig:Diagrams}, we present the probability distributions that the same sample of stars belong to the thin disk (bottom left), the thick disk (bottom in the middle), and to the halo (bottom right). Most of the stars are likely members of the thin disk. Only 10$\%$ of these stars have a probability $>90\%$ (as TOI-5076b) of belonging to the thick disk, while all of them have a negligible probability of belonging to the halo.
      
The planet's equilibrium temperature ($\rm T_{eq}$) can be estimated with

\begin{equation}
T\rm_{eq} = T_{\ast}\sqrt{\frac{R_{\ast}}{2\,a}}\Big(1-A\Big)^{1/4}
,\end{equation}

\noindent
where $a$ is the orbital semi-major axis, $R_{\ast}$ the stellar radius,  $T_{\ast}$ the host star's effective temperature, and $A$ the Bond albedo\footnote{We assumed here full heat redistribution and zero Bond albedo.}. We obtained  
$T\rm_{eq}=(615\pm20)$ K.

Assuming an H$_2$-dominated solar-abundance atmosphere, the scale height of the planetary atmosphere is H=$\rm\frac{k_b\,T_{eq}}{\rm\mu\,g}$, where $k_b$ is the Boltzmann constant, $T\rm_{eq}$ is the planet's equilibrium temperature, $\mu=2.3$ amu is the mean molecular weight, and $g$ is the gravity. For TOI-5076b we have H = 142 km, while the amplitude of spectral features in transmission for a clear atmosphere is $\rm\sim4pH/R\rm_{\ast}=39 $~ppm \citep{kreidberg2018}, where p is the radius ratio and R$\rm_{\ast}$ is the radius of the star.

We also calculated the transmission spectroscopy metric 
\citep[TSM; ][]{kempton2018} to understand the suitability of this target for transmission spectroscopy using the James Webb Space Telescope (JWST). The TSM parameter provides an estimate of the S/N reachable in a 10 hr observing program using JWST/NIRISS, and it is defined as

\begin{equation}
\mbox{TSM} = S \times \frac{R_p^3T_{eq}}{M_pR_{\ast}^2} \times 10^{-m\rm_J/5}
,\end{equation}

\noindent
where S is a normalization constant to match the more detailed work of
\citet{louie2018}, $R_p$ is the radius of the planet in units of Earth radii, $M_p$ is the mass of the planet in units of Earth masses, $R_{\ast}$ is the radius of the host star in units of solar radii, $m\rm_J$ is the apparent magnitude of the host star in the J band, and T$\rm_{eq}$ is the equilibrium temperature expressed in Kelvin. From Table~1 of \citet{kempton2018}, we chose S=1.28 considering the radius of TOI-5076b. We found a value of TSM = $29$. 
This places TOI-5076b in the fourth quartile of the TSM values of the statistical sample of planets simulated by \citet{kempton2018} and is below the suggested cut-off 
value (TSM=84) for follow-up efforts. 

Therefore, the atmospheric characterization of TOI-5076b can be considered challenging in comparison with other more favorable targets. We could also discuss tentative trends identified in the small sample of warm Neptunes studied so far \citep{crossfield2017}, which suggests that both T$\rm_{eq}$ and f$\rm_{HHe}$ (H and He mass fractions, which can be traced by r$\rm_p$) play a key role in setting the amplitude of the H$_2$O absorption feature at 1.4$\mu$m in transmission spectroscopy. The relatively low equilibrium temperature\footnote{To be consistent with \citet{crossfield2017}, we assumed full heat redistribution and Bond albedo $A$=0.2 here.} (T$\rm_{eq}=581$ K) of TOI-5076b would fall below the threshold identified by \citet{crossfield2017} for the formation of hazes in the atmosphere of warm Neptunes (T$\rm_{eq}=850$ K). Therefore, most or all transmission features should be blocked in TOI-5076b. While the density of TOI-5076b is compatible with the presence of an H/He envelope (see Fig.~\ref{fig:RhopRp}), its relatively large mass (Fig.~\ref{fig:mass_radius}) could suggest a massive rocky core and/or a metal-rich atmosphere (as we note above), which would also decrease the amplitude of atmospheric features in transmission.



\section{Conclusions}
\label{sec:conclusions}

In this work, we report the confirmation of a new transiting exoplanet orbiting the star TOI-5076. Using HARPS-N spectroscopy, we determined that the host star is a metal-rich K2V dwarf located about 82 pc from the Sun with a radius of R$_{\star}$=(0.78$\pm$0.01) R$_{\odot}$ and a mass of M$_{\star}$=(0.80$\pm$0.07) M$_{\odot}$. It forms a common proper motion pair with an M-dwarf companion star located at a projected separation of 2178 au. The chemical analysis of the host star and the Galactic-space velocities indicate that TOI-5076 belongs to the old population of thin-to-thick-disk transition stars. 

The transiting planet is a warm (T$\rm_{eq}$=614$\pm$20 K) sub-Neptune with a mass of m$\rm_p=$(16$\pm$2) M$\rm_{\oplus}$, and a radius of r$\rm_p=$(3.2$\pm$0.1)~R$\rm_{\oplus}$ yielding a density of $\rho_p$=(2.8$\pm$0.5) g cm$^{-3}$. Its orbital period is ~23.445 days.

The density of TOI-5076b suggests the presence of a large fraction by volume of volatiles overlying a massive core. We find that a circular orbit solution is marginally favored with respect to an eccentric orbit solution for TOI-5076b with the present data. 
In the case of the eccentric orbit solution, we obtained e=(0.20$\pm$0.09).
The analysis of the photometric measurements and radial velocities after subtraction of the best-fit model do not show the presence of any additional planet in the system.

\begin{acknowledgements}
 Based on observations made with the Italian {\it Telescopio Nazionale
Galileo} (TNG) operated by the {\it Fundaci\'on Galileo Galilei} (FGG) of the {\it Istituto Nazionale di Astrofisica} (INAF) at the {\it  Observatorio del Roque de los Muchachos} (La Palma, Canary Islands, Spain). 
This paper made use of data collected by the TESS mission and are publicly available from the Mikulski Archive for Space Telescopes (MAST) operated by the Space Telescope Science Institute (STScI). Funding for the TESS mission is provided by NASA’s Science Mission Directorate. We acknowledge the use of public TESS data from pipelines at the TESS Science Office and at the TESS Science Processing Operations Center. Resources supporting this work were provided by the NASA High-End Computing (HEC) Program through the NASA Advanced Supercomputing (NAS) Division at Ames Research Center for the production of the SPOC data products.
This work made use of \texttt{tpfplotter} by J. Lillo-Box (publicly available in \url{http://www.github.com/jlillo/tpfplotter}), which also made use of the python packages \texttt{astropy}, \texttt{lightkurve}, \texttt{matplotlib} and \texttt{numpy}. This research has made use of the Exoplanet Follow-up Observation Program (ExoFOP; DOI: 10.26134/ExoFOP5) website, which is operated by the California Institute of Technology, under contract with the National Aeronautics and Space Administration under the Exoplanet Exploration Program. Some of the observations in the paper made use of the NN-EXPLORE Exoplanet and Stellar Speckle Imager (NESSI). NESSI was funded by the NASA Exoplanet Exploration Program and the NASA Ames Research Center. NESSI was built at the Ames Research Center by Steve B. Howell, Nic Scott, Elliott P. Horch, and Emmett Quigley. B.S.S. and I.A.S. acknowledge the support of M.V. Lomonosov Moscow State University Program of Development.
TZi acknowledges NVIDIA Academic Hardware Grant Program for the use of the Titan V GPU card and the support by the CHEOPS ASI-INAF agreement n. 2019-29-HH.0 and the Italian MUR Departments of Excellence grant 2023-2027 “Quantum Frontiers”.
\end{acknowledgements}

%
   \bibliographystyle{aa} 
   \bibliography{example} 
%

\end{document}